\documentclass[a4paper,12pt]{article}

\pdfoutput=1 

\usepackage[hmargin=.71in,vmargin=1.1in]{geometry}
\usepackage{indentfirst}
\usepackage{amsfonts}
\usepackage{mathrsfs}
\usepackage{amsmath}
\usepackage{mathabx}
\usepackage{amssymb}
\usepackage{authblk}
\usepackage{cite}
\usepackage{xcolor}
\usepackage{mathtools}
\usepackage{tensor}
\usepackage{physics}
\usepackage{graphicx}
\usepackage{bm}
\usepackage{upgreek}
\usepackage{braket}
\usepackage{color,soul}
\usepackage{csquotes}
\usepackage{caption}
\usepackage{subcaption}
\usepackage{slashed}
\usepackage{mathtools}
\usepackage{multirow}

\usepackage[bookmarksnumbered=true,bookmarksopen=true]{hyperref}
 \hypersetup{colorlinks,
             linkcolor=[rgb]{0,0.3,0.6}, 
             citecolor=[rgb]{0,0.3,0.6}, 
             urlcolor=[rgb]{0,0.3,0.6}}

\newcommand\me{\mathrm{e}}

\newcommand\pp{\uppi}
\newcommand{\dif}{\mathrm{d}}

\usepackage{booktabs}
\usepackage{multirow}

\usepackage{tikz}
\usetikzlibrary{matrix}

\begin{document}

\title{\Large\textbf{Quasinormal modes of regular black holes surrounded by skewed dark matter distributions}}

\author[a]{Chen Lan\thanks{stlanchen@126.com}}
\author[b]{Ying-Lei Tian\thanks{yl.tian@mail.nankai.edu.cn}}
\author[c]{Hao Yang\thanks{hyang@ucas.ac.cn}}
\author[b]{Zhen-Xiao Zhang\thanks{zx.zhang@mail.nankai.edu.cn}}
\author[b]{Yan-Gang Miao\thanks{Corresponding author: miaoyg@nankai.edu.cn}}

\affil[a]{\normalsize{\em Department of Physics, Yantai University, 30 Qingquan Road, Yantai 264005, China}}
\affil[b]{\normalsize{\em School of Physics, Nankai University, 94 Weijin Road, Tianjin 300071, China}}
\affil[c]{\normalsize{\em School of Fundamental Physics and Mathematical Sciences, Hangzhou Institute for Advanced Study, UCAS, Hangzhou 310024, China}}

\date{ }

\maketitle

\begin{abstract}

Regular black holes, nonsingular solutions to gravitational collapse with quantum corrections, 
offer a compelling alternative to classical black holes with curvature singularities. 
In this work, we investigate how the presence of skewed dark matter distributions outside 
the innermost stable circular orbit of regular black holes modifies the gravitational wave signals emitted by such objects. 
Rather than introducing corrections directly into an effective potential, 
we model the influence of dark matter through metric corrections, allowing a full control over 
the spatial distribution and abundance of dark matter. We demonstrate that a skewed normal profile 
generically introduces shallow potential wells or secondary barriers in the effective potential of perturbation equations, 
depending sensitively on the type of perturbations: scalar, spinor, or tensor. 
These modifications lead to distinctive quasinormal mode features, including long-lived modes, echo effects,  
and in some cases, altered stability behaviors. Notably, the axial and polar sectors of tensor field perturbations 
respond asymmetrically to identical dark matter profiles, revealing a deeper structural distinction in their perturbation dynamics. 
These results provide a theoretical framework for probing regular black holes in the dark matter environment through gravitational wave observations.

\end{abstract}

\tableofcontents

\section{Introduction}

Dark matter is one of the most mysterious components of the universe~\cite{Bertone:2016nfn,Planck:2018nkj}. 
According to the standard $\Lambda$CDM cosmological model, it accounts approximately for $27\%$ of the total energy density of the universe \cite{DiValentino:2021izs,Perivolaropoulos:2021jda}. 
Owing to its defining property --- it interacts only via gravitational but not via any other fundamental forces, dark matter is expected to cluster more densely in the regions of strong gravity, such as in the vicinity of black holes \cite{Xu:2018wow,Zhang:2020mxi,Narzilloev:2020qtd}. 
This dense accumulation of dark matter can potentially leave observable signatures \cite{Sadeghian:2013laa,Ferrer:2017xwm,Cole:2022yzw,Maeda:2024tsg,Cardoso:2021wlq}, including modifications to black hole shadows \cite{Jusufi:2019ltj,Jusufi:2019nrn,Konoplya:2019sns,Saurabh:2020zqg,Jusufi:2020cpn} and gravitational wave signals \cite{Eda:2014kra,Coogan:2021uqv,Kavanagh:2020cfn,Figueiredo:2023gas}.
In this work, we investigate how the presence of dark matter influences gravitational wave signals in the vicinity of a particular class of black holes,  called regular or nonsingular black holes, where the dark matter is distributed with a skewed normal distribution outside the innermost stable circular orbit (ISCO) of regular black holes.
In other words, the distribution is asymmetric and has a higher density of dark matter closer to but a lower density away from  regular black holes.

Regular black holes have emerged as one of the most active research topics in the contemporary black hole physics \cite{Bonanno:2023rzk,Bueno:2024dgm,Bueno:2024eig,Bronnikov:2024izh,Bueno:2024zsx,Carballo-Rubio:2022nuj}. The term `regular' refers to black holes that are free of spacetime 
singularities,
i.e., that are typically characterized by geodesic completeness as per the singularity theorems. 
Moreover, this term is often accompanied by curvature finiteness in spacetime.\footnote{However, these two definitions are not always consistent. For a detailed discussion of this issue, we refer the reader to Ref.\ \cite{Zhang:2025nim}.}
The absence of a singularity is typically attributed to the presence of a core at the center of regular black holes that violates the strong energy condition, thereby evading the Penrose singularity theorem and effectively removing the singularity~\cite{Bambi:2023try,Lan:2023cvz}.

In this work, we focus exclusively on a class of recently proposed regular black holes that arises from the gravitational collapse of matter \cite{Bonanno:2023rzk,Cipriani:2024nhx,Vertogradov:2025yto,Vertogradov:2025snh}, supplemented by quantum corrections, such as those predicted by asymptotic safety \cite{Bonanno:2023rzk} and loop quantum gravity theories \cite{Cipriani:2024nhx}.
These models are of particular interest not only because they provide a concrete mechanism for the formation of regular black holes, but more importantly, because their construction implies that the spacetime outside horizons can carry imprints of their interior structure. In this class of regular black holes, the interior spacetime is first determined by the collapse of matter, including possible quantum corrections, and the exterior metric is then derived via appropriate matching conditions \cite{Israel:1966rt,Barrabes:1991ng,Fayos:1996gw}. As a result, such models offer a potential window into probing the otherwise inaccessible spacetime inside event horizons.

Moreover, we employ  the quasinormal mode (QNM) analysis \cite{Berti:2025hly} to study the impact of dark matter on the gravitational waves emitted by regular black holes. QNMs refer to damped oscillation modes that characterize the  waveform of ringdown phases following binary black hole mergers \cite{Konoplya:2011qq}. The QNM analysis is widely used in the current study of regular black holes\cite{Konoplya:2022hll,Konoplya:2023aph,Konoplya:2023ahd,Konoplya:2025hgp,Lan:2023vaa}, most of which focus, however, on vacuum solutions, leaving matter effects on singular black holes.
The recent research shows that the presence of matter modifies the effective potential in the perturbation equations \cite{Li:2019kwa,Cheung:2021bol,Huang:2021qwe,Berti:2022xfj,Cao:2024sot,Mai:2025cva}, introducing a second potential barrier or well, which in turn affects the stability of perturbations.
Our work extends the research by incorporating a skewed dark matter distribution around regular black holes, revealing novel features, such as long-lived modes and enhanced polar perturbation sensitivity, which would probably inform future gravitational wave observations.

There are two formally distinct methods to modify the effective potential of perturbation equations. The first method involves directly adding a correction term to the effective potential function of tortoise coordinates, which allows for a precise control over the position of corrections relative to potential barriers. However, it is accompanied by certain challenges.
The first one is the issue of physical interpretations: What kind of matter distributions around black holes would correspond to such a corrected potential? 
Since this method essentially infers the mass distribution from the modified effective potential, it can lead to the so-called “negative density” problem, where the inferred density becomes negative in certain spatial regions.
The second challenge comes from a technical limitation: The correction term depends on the tortoise coordinate, which generally lacks an analytic form except for a few simple models. As a result, numerical calculations are required to implement the modification in most cases.

The second method involves modifying the black hole metric itself by introducing a correction to the physical mass. This strategy offers a better control over the shape and abundance of the corrected mass distribution. However, it has the drawback that it does not allow a precise control over the exact central location of matter distributions. Even a small modification to the metric can shift the positions of key features, such as the event horizon, photon sphere, and ISCO.
In this work, we adopt the second method because the exact location of matter distributions is not the primary concern for our purpose. Instead, our main objective is to investigate how the matter distributed outside the ISCO influences the gravitational waveform. Thus, it suffices to ensure that the matter is located beyond the ISCO.

The structure of this paper is as follows. In Sec.~\ref{sec:original}, we begin by analyzing the ringdown waveform of the model without dark matter~\cite{Bonanno:2023rzk}. When the parameter $\xi$ approaches zero, the model reduces exactly to the Schwarzschild black hole, meaning that $\xi$ characterizes the deviation from Schwarzschild geometry. Our analysis reveals that the ringdown waveform exhibits distinct features from those of the Schwarzschild black hole, particularly during the initial phase.\footnote{Our classification of waveforms follows the conventions established in Refs.\ \cite{Konoplya:2011qq,Burgess:2023pny} (see Fig.~3 in either reference). The waveform is typically divided into three distinct stages: The initial transient phase, the ringdown phase, and the late-time tail phase.}

In Sec.\ \ref{sec:testfield}, we introduce dark matter with a skewed normal distribution and derive its impact on the effective potential of perturbation equations. For test field perturbations, we find that the presence of dark matter introduces an additional complex structure into the effective potential. Furthermore, this modification leads to a “long-lived” behavior in the waveform, i.e., the oscillations become significantly prolonged and the oscillation frequencies gradually decrease.

Section \ref{sec:tensorpert} focuses on tensor field perturbations in the presence of dark matter. It is important to note that we neglect the backreaction of the stress-energy tensor on the perturbation equations, which is a reasonable approximation. Our main goal here is to investigate whether the same form of dark matter distributions causes different modifications to the effective potentials of axial and polar tensor field perturbations. 
Our results clearly indicate that, despite adopting the same dark matter profile, 
the polar perturbations exhibit significantly enhanced sensitivity compared to their axial counterparts: Even modest dark matter abundances can generate spectral peaks that substantially exceed the amplitude of the dominant mode without dark matter. This differential behavior manifests directly in the corresponding ringdown waveform, providing a potentially observable signature of dark matter effects on gravitational wave signals.

Section \ref{sec:dis} presents a general discussion and outlook. At last, we explore how the same dark matter distribution affects the ringdown waveform of Schwarzschild black holes in App.~\ref{app:schwarzschild}.

\section{Quasinormal modes without dark matter}
\label{sec:original}
In this section, we revisit the QNMs of a regular black hole derived from the framework of asymptotically safe gravity. Unlike Ref.~\cite{Stashko:2024wuq}, which focused on the QNM spectrum and gray-body factors, we make our analysis from the perspective of waveform evolution. 
This distinction ensures that our investigation is not a repetition of previous calculations.
Our primary goal here is to establish a complete baseline for comparison, particularly with the dark-matter-modified model introduced in the following sections.

The black hole model under consideration is formed via dust collapse, with quantum corrections inspired by asymptotic safety. Its exterior spacetime metric takes~\cite{Bonanno:2023rzk} the form,
\begin{equation}
\label{eq:metric_orig}
\dif s^2 = -f(r) \dif t^2 + f^{-1}(r) \dif r^2 + r^2 \dif \Omega^2,
\end{equation}
and the shape function $f(r)$ is given by
\begin{equation}
\label{eq:metric_shape}
f(r) = 1 - \frac{r^2 }{3 \xi}\ln \left(1+ \frac{6 M \xi }{r^3}  \right),
\end{equation}
where $M$ is the mass and the parameter $\xi$ is supposed to be determined by observations.
Here, $f(r)$ converts to that of Schwarzschild black holes as $\xi\to0$, i.e., $\xi$ determines how much the model deviates from Schwarzschild black holes.
According to the shape function, the horizon can be determined by
\begin{equation}
  \xi = -\frac{r_{\rm H}^3}{6 M}-\frac{r_{\rm H}^2}{3}  
  W_{-1}\left(-\frac{r_{\rm H}}{2 M}\me^{-\frac{r_{\rm H}}{2 M}}\right),
\end{equation}
where $W_{-1}$ is a branch of the Lambert W function, and the other branch is $W_0$.

In this model, the parameter $\xi$ possesses a critical value, denoted by $\xi_c=2y^2(3+2y)M^2/3$ with $y=W_0(-3/(2\me^{3/2}))$. When $\xi = \xi_c$, the spacetime admits a single degenerate horizon $r^{\rm ext}_{\rm H}=-2y M$; for $\xi > \xi_c$, two distinct horizons exist, while for $\xi < \xi_c$, the spacetime has no horizons.

Moreover, at $\xi = \xi_c$ 
the radius of the photon sphere is
\begin{equation}
    r_{\rm ps}=3 M \left(\frac{y}{3}+\frac{1}{2}+\frac{1}{2} \sqrt{1-\frac{4}{3} y (y+1)}\right)\approx 2.59 M,
\end{equation}
while the radius of the ISCO, defined by the conditions $V_{\rm eff}' =0= V_{\rm eff}'' $, gives
$r_{\rm ISCO} \approx 5.51  M$.
Here, $V_{\rm eff}=f(r) \left(\frac{L^2}{r^2}+1\right)$ refers to the effective potential for a massive test particle moving along geodesics \cite{Chandrasekhar:1985kt,Guo:2025scs}, where $L$ denotes the angular momentum of the test particle. 
To distinguish it from the effective potential in the perturbation equations discussed later, we explicitly denote the former with the subscript ``eff'' and use the calligraphic symbol $\mathcal{V}$ for the latter without subscripts. 
We mention in advance that the location of the ISCO will play a key role in our modeling of the dark matter distribution in the subsequent sections.

\begin{figure}[!ht]
	\centering
		\includegraphics[width=.6\textwidth]{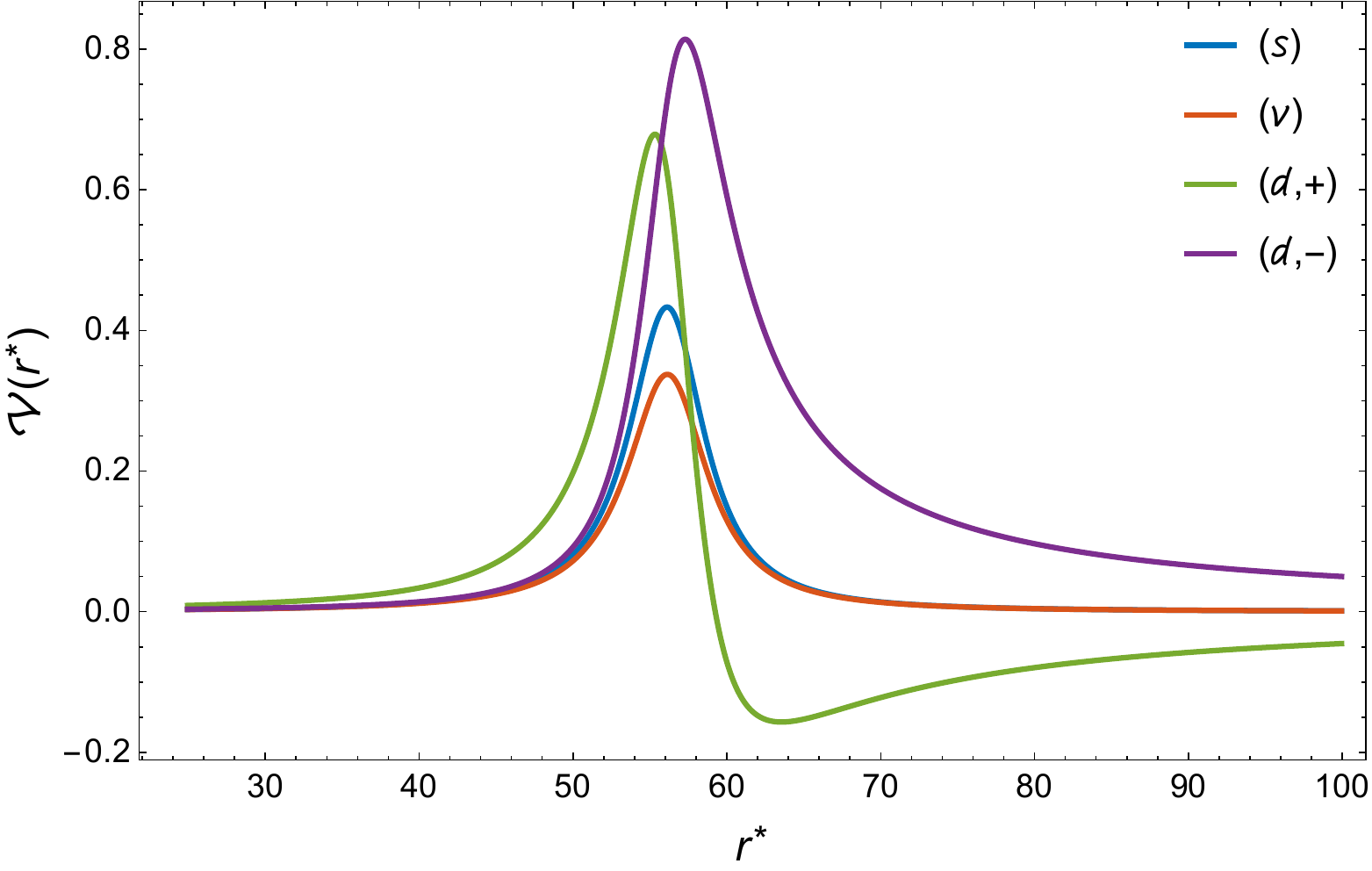}

	\captionsetup{width=.9\textwidth}
	\caption{
		Effective potentials of different test fields with $\ell=1$.}
	\label{fig:Veff_orig}
\end{figure}

We now examine the perturbations of three types of test fields in the single-horizon background. 
The perturbations of massless test fields in this background satisfy master equations of the form $\frac{d^2 \Psi}{dr_*^2} + (\omega^2 - V(r)) \Psi = 0$, where $r_*$ is the tortoise coordinate, $\omega$ is the mode frequency, and $\Psi$ is the radial wave function \cite{Chandrasekhar:1985kt,Pani:2013pma}. 
These equations are derived from the Klein-Gordon equation for scalar fields, the Maxwell equations for vector fields, and the Dirac equation for spinor fields, after variable separation and appropriate rescaling. The corresponding effective potentials are~\cite{Zinhailo:2019rwd,Stashko:2024wuq}
\begin{subequations}
\begin{equation}
\mathcal{V}^{(s)}(r) =
f(r) \left[
\frac{\ell(\ell+1)}{r^2}
+\frac{1}{r} \frac{\dif f(r)}{\dif r}
\right],
\end{equation}
\begin{equation}
\mathcal{V}^{(v)}(r) =
f(r) \left[
\frac{\ell(\ell+1)}{r^2}    
\right],
\end{equation}
\begin{equation}
\mathcal{V}^{(d,\pm)}(r) =
f(r) \frac{\ell+1}{r}\left[
\frac{\ell+1}{r}
\mp \sqrt{f(r)}
\pm \frac{\dif }{\dif r} \sqrt{f(r)}
\right],   
\end{equation}
\end{subequations}
where the superscripts $(s)$, $ (v)$, and $(d,\pm)$ on the potential $\mathcal{V}(r)$ label scalar, vector, and spinor field perturbations, respectively. The tortoise coordinate is defined via $r^* = \int \frac{\dif r}{f(r)}$. The corresponding effective potentials\footnote{It is important to note that for positive-parity spinor field perturbations, if $\ell = 0$, the effective potential takes the form of a pure potential well, whereas if $\ell > 1$, it becomes a pure potential barrier. A similar behavior is also observed in the case of axial tensor field perturbations.} are illustrated in Fig.\ \ref{fig:Veff_orig}.

As shown in Fig.\ \ref{fig:Veff_orig}, the potentials exhibit standard bell-shaped profiles for the scalar $(s)$, vector $(v)$, and negative-parity spinor $(d,-)$ field perturbations. However, the positive-parity spinor field perturbation $(d,+)$ displays a qualitatively different structure, presenting a potential well located to the right of the main barrier.
According to our previous studies \cite{Zhang:2025dzt,Lan:2025cvs}, this kind of effective potential structures can give rise to unstable QNMs, meaning that the associated perturbation waveform grows unbounded over time. The onset of such instability is determined by the depth of the potential well, particularly the ratio between the peak of the potential barrier and the depth of the well.

\begin{figure}[!ht]
	\centering
	\begin{subfigure}[b]{0.45\textwidth}
		\centering
		\includegraphics[width=\textwidth]{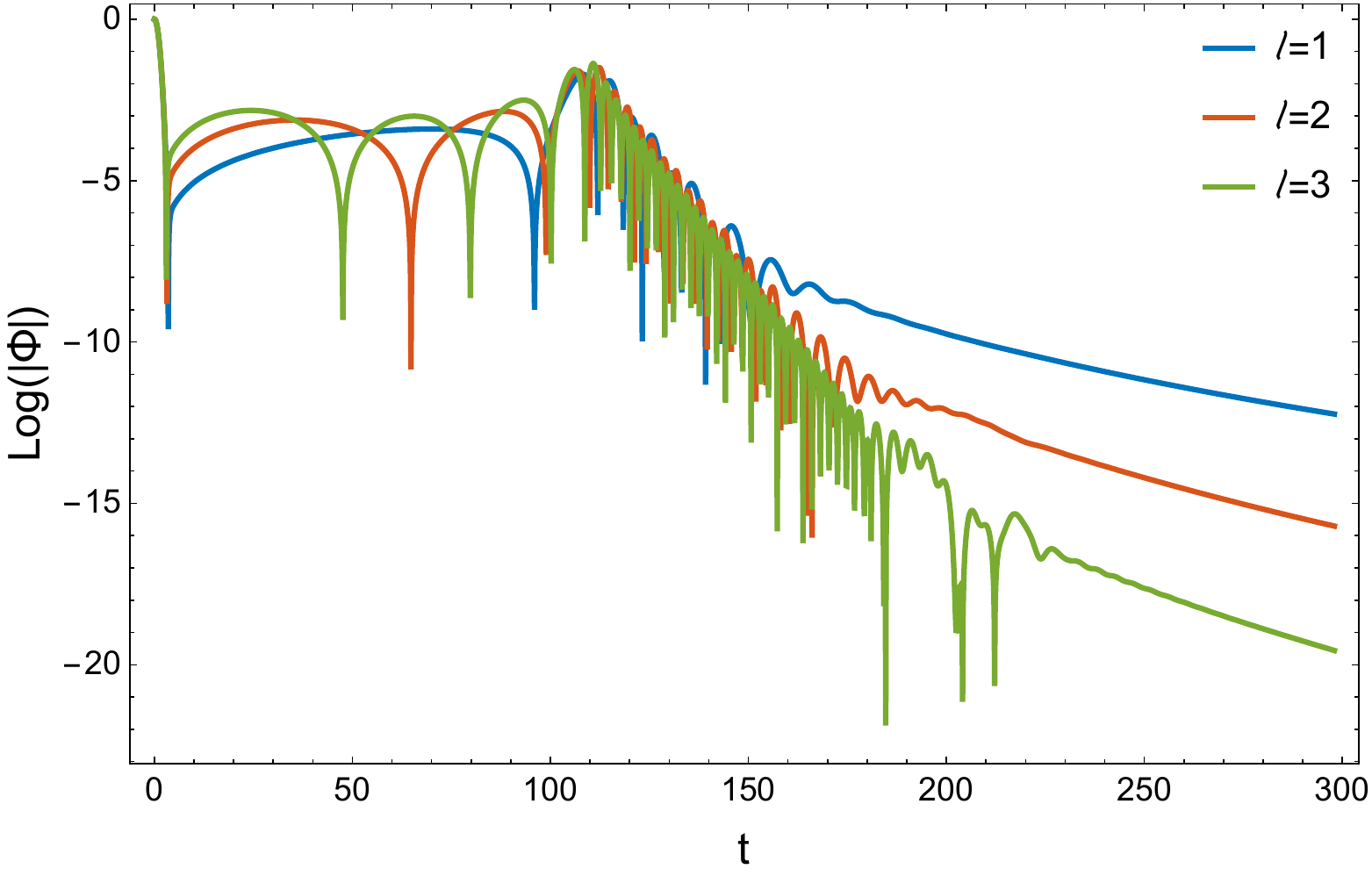}
		\caption{Scalar $(s)$}
		\label{fig:s_wf_orig}
	\end{subfigure}
	\begin{subfigure}[b]{0.45\textwidth}
		\centering
		\includegraphics[width=\textwidth]{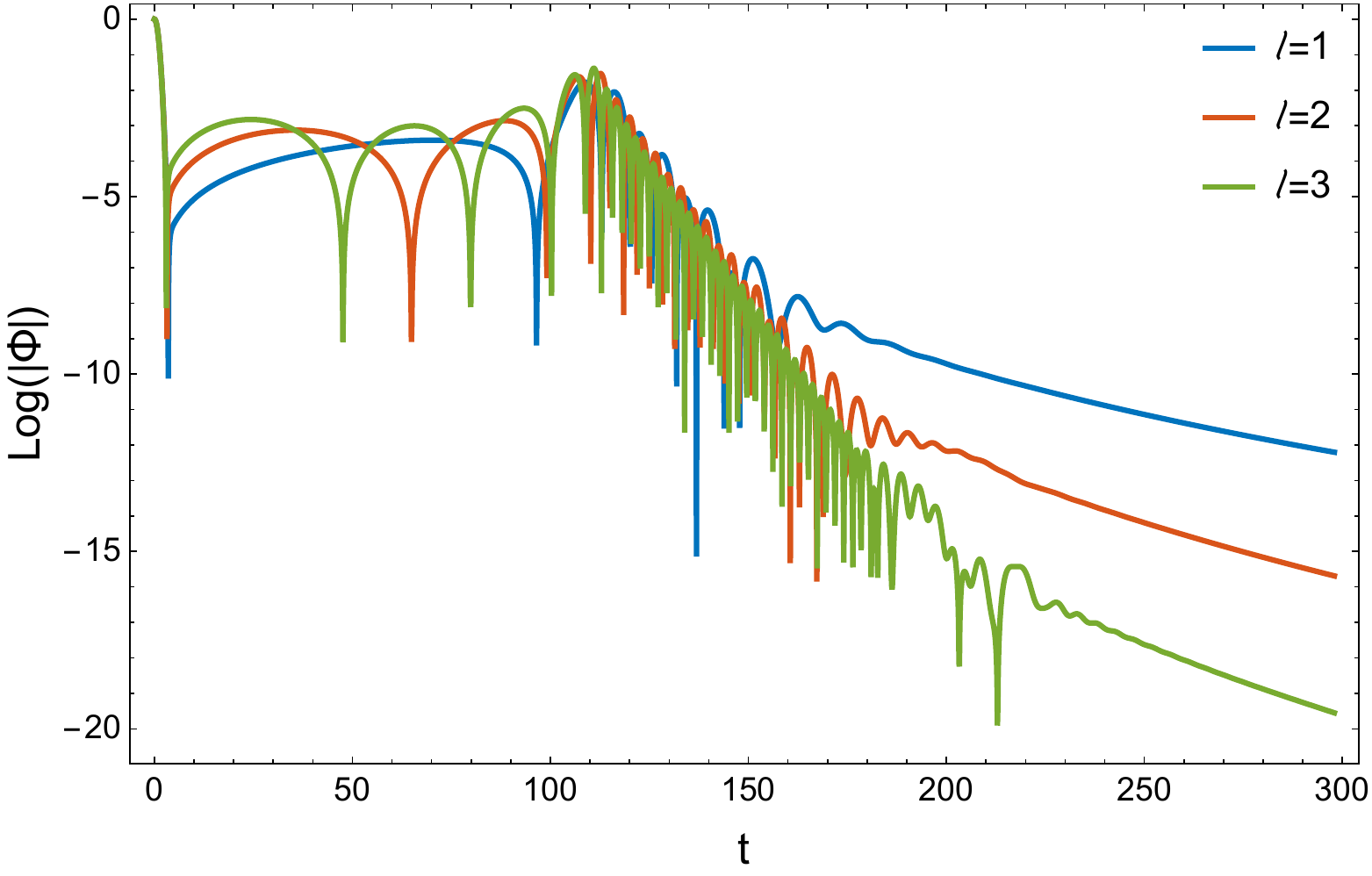}
		\caption{Vector $(v)$}
		\label{fig:v_wf_orig}
	\end{subfigure}
	\begin{subfigure}[b]{0.45\textwidth}
		\centering
		\includegraphics[width=\textwidth]{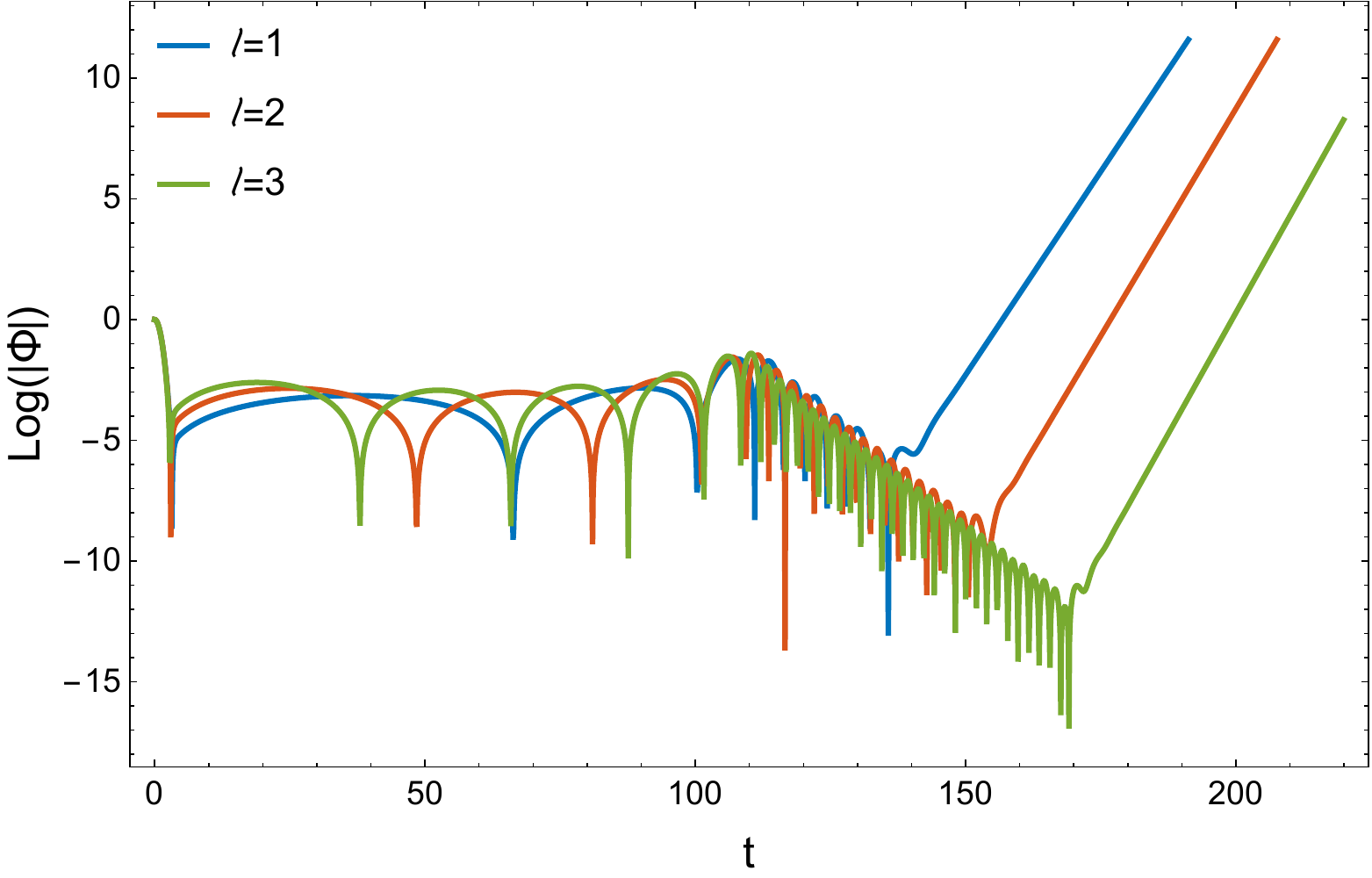}
		\caption{Spinor $(d,+)$}
		\label{fig:sp_wf_orig}
	\end{subfigure}
	\begin{subfigure}[b]{0.45\textwidth}
		\centering
		\includegraphics[width=\textwidth]{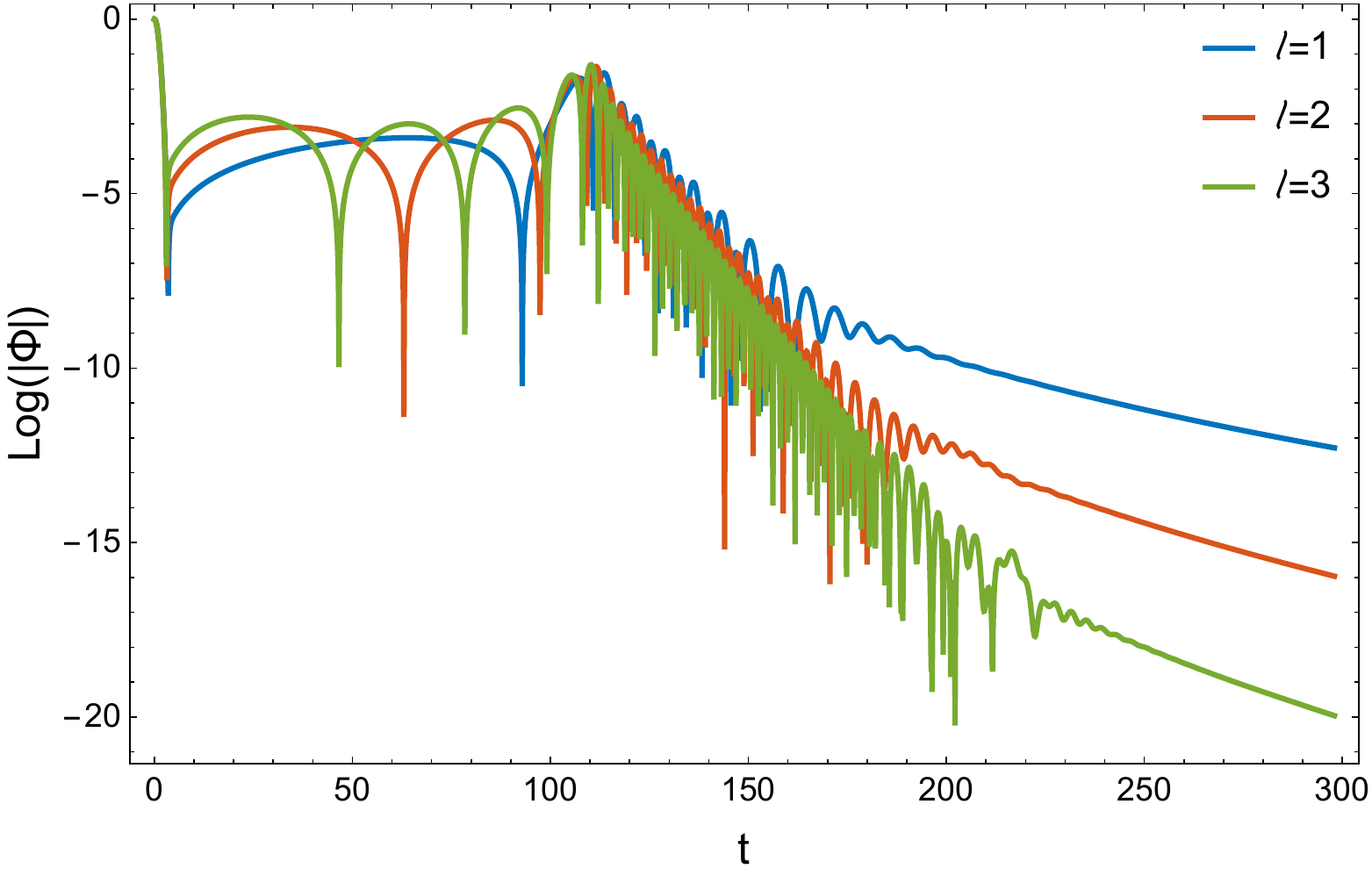}
		\caption{Spinor $(d,-)$}
		\label{fig:sm_wf_orig}
	\end{subfigure}
	\captionsetup{width=.9\textwidth}
	\caption{Ringdown waveform of scalar, vector and spinor field perturbations.}
	\label{fig:wf_orig}
\end{figure}

The ringdown waveform is shown in Fig.\ \ref{fig:wf_orig}, 
where the initial perturbation is modeled as a Gaussian wave packet.\footnote{The ringdown waveform is computed using a null coordinate $(u,v)$ grid discretization of the wave equation with a four-point finite difference stencil, achieving $O(h^4)$ local accuracy and systematic integration along null characteristics \cite{Gundlach:1993tp,Zhang:2025dzt}.}
Due to the structural similarity of the effective potentials, panels \ref{fig:s_wf_orig}, \ref{fig:v_wf_orig}, and \ref{fig:sm_wf_orig} exhibit similar waveform features: The signal initially displays damped oscillations, and then gradually diminishes into a monotonic decay.
As the angular quantum number $\ell$ increases, both the oscillation frequency and the decay rate increase, corresponding  to a larger real and minus imaginary parts of QNM frequencies, respectively, which results in a longer-lasting oscillation and a faster subsequent damping.

Panel \ref{fig:sp_wf_orig}, as predicted from the structure of the effective potential, exhibits a qualitatively different behavior: The late-time tail waveform grows exponentially over time, signaling an instability. This divergence is consistent with the presence of a potential well beyond the main barrier.
Moreover, as $\ell$ increases from $1$ to $2$ and to $3$, the absolute value of barrier peak-to-well depth ratios grows monotonically from $3.41$ to $6.28$ and to $10.54$. This indicates that the influence of the potential well weakens progressively, and accordingly, the onset of the late-time tail divergence is delayed.

\section{Effect of dark matter on quasinormal modes}
\label{sec:testfield}

In this section, we  introduce dark matter into the model depicted by Eqs.~\eqref{eq:metric_orig} and \eqref{eq:metric_shape}.
From a physical perspective, we consider the scenario in which the dark matter is distributed outside the ISCO. 
Its mass distribution follows a positively skewed bell-shaped profile, resembling the matter configuration in a standard accretion disk.

\begin{figure}[!ht]
	\centering
	\includegraphics[width=.35\textwidth]{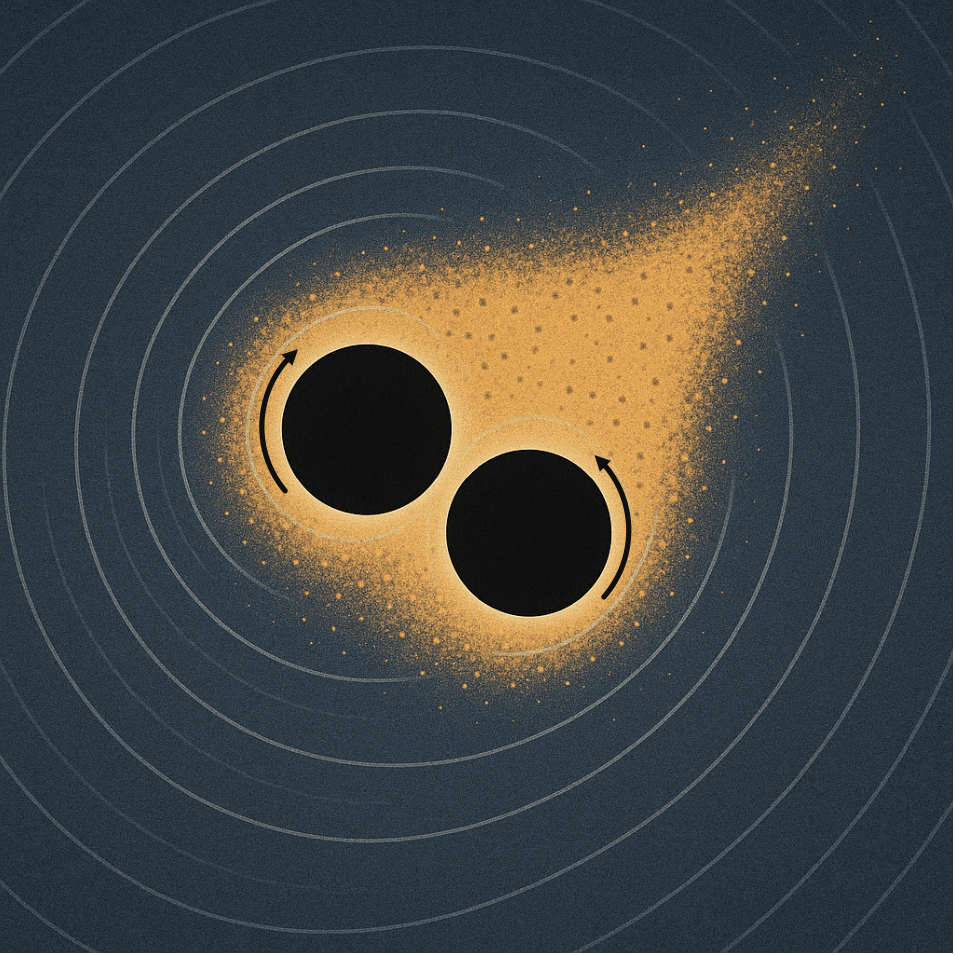}
	\captionsetup{width=.9\textwidth}
	\caption{Illustration: Gravitational wave emission from a binary black hole system immerged in a dark matter environment.
	}
	\label{fig:BH_DM}
\end{figure}

Figure \ref{fig:BH_DM} presents a schematic illustration of a binary black hole merger occurring in a dark matter environment. The two black discs represent rotating black holes, where the arrows indicate their spin directions. The ochre-colored background denotes the surrounding dark matter halo that is undergoing the accretion around the merging black holes during the coalescence process. The concentric gray ripples depict the emitted gravitational wavefronts. It is important to emphasize that this figure serves purely as a conceptual visualization of the underlying physical mechanism explored in this work, but does not represent the outcome of a detailed numerical simulation or an actual modeling result.

Our assumption of a quasi-stationary dark matter distribution simplifies the analysis but neglects dynamical effects such as accretion, where the dynamical effects shift the ISCO and alter the ringdown waveform. Additionally, our focus on non-rotating black holes limits applicability to astrophysical scenarios where rotation is significant. Future work will use numerical relativity to model time-dependent dark matter distributions and extend our analysis to rotating black holes using the Newman-Janis algorithm.

\subsection{Introducing dark matter}

We begin our analysis with the setup of dark matter, where the positively skewed function describing the 
density distribution is given by
\begin{equation}
\label{eq:dm_mass}
\rho_{\rm DM}(r) = \rho_0 
\me^{-\frac{\left(r-r_0\right)^2}{2 \sigma ^2}} \left[1 +\text{erf}\left(-\frac{\alpha  \left(r-r_0\right)}{\sqrt{2} \sigma }\right)\right],\qquad 
r\gtrsim r_{\rm ISCO},
\end{equation}
where $\rho_0$ is a parameter with density dimension, 
$r_0\gtrsim r_{\rm ISCO}$,\footnote{Actually, the presence of a radial-dependent mass correction inevitably shifts the location of the ISCO in the modified model relative to that in the unmodified case.}
$\sigma$ is the width, and $\alpha>0$ is  shape parameter.
This density profile is a phenomenological choice inspired by skewed distributions in accretion
disks and dark matter halos, where the asymmetry arises from gravitational clustering \cite{Binney:2008gds,Navarro:2003ew,Springel:2008cc}. It resembles
the matter configuration in standard thin-disk models but with skewness to reflect higher densities near
the black hole.
In our analysis, the peak position $r_0$ of the dark matter distribution is chosen to lie outside the ISCO, motivated by considerations of environmental stability. If dark matter were distributed inside the ISCO, it would continuously fall into the merged black hole, leading to a time-dependent evolution of the spacetime and corresponding modifications to the effective potential. Such dynamical scenarios are inherently more complex and are not addressed in the present work. In contrast, when $r_0 > r_{\rm ISCO}$, the dark matter configuration can be reasonably treated as quasi-stationary, resulting in a static modification to the effective potential.

The skewed normal distribution implies that the dark matter density is high near black holes, but low away from black holes. This assumption comes from the strong gravity of black holes, which is fully consistent with physically realistic scenarios.
It is worth emphasizing that there exist many possible implementations of skewed normal distributions. In this work, we adopt a representative form, namely, a modified Gaussian profile with skewness, to illustrate the essential physical features without loss of generality.
In addition, the mass density profile introduced in Eq.~\eqref{eq:dm_mass} provides an opportunity for comparison with the well-established Einasto profile~\cite{Navarro:2003ew,Graham:2005xx}, which is commonly used in modeling dark matter halos,
\begin{equation}
\rho_{\rm Einasto}(r) = \rho_{-2} \exp\left[
-\frac{2}{\beta} \left(
\left( \frac{r}{r_{-2}} \right)^\beta - 1
\right)
\right],
\end{equation}
where $r_{-2}$ is the radius at which the logarithmic slope of the profile equals $\dif \ln \rho/\dif \ln r=-2$, and $\beta$ controls the curvature of the density profile.

If the parameters $(\rho_{-2}, \beta)$ of the Einasto profile can be determined from observational data, 
we may then calibrate the parameters of our dark matter profile $\rho_{\rm DM}$ accordingly at infinity. In other words, under the weak field approximation, the two density distributions should be roughly equal.
For instance, we can set $r_0 \approx r_{-2}$ and $\rho_0\approx \rho_{-2}$ in $\rho_{\rm DM}$, and considering the difference in the density profiles directly at infinity, we have the leading term,
\begin{equation}
\ln(\rho_{\rm DM}) - \ln(\rho_{\rm Einasto}) \approx \frac{2 r^{\beta } }{\beta r_{-2}^{\beta }}-\frac{\left(\alpha ^2+1\right) r^2}{2 \sigma ^2},
\end{equation}
which yields the following estimate for the parameters,
\begin{equation}
\beta\approx 2,\qquad
\frac{\sigma }{\sqrt{\alpha ^2+1}}\approx\frac{r_{-2}}{\sqrt{2}}.
\end{equation}
These approximations allow us to relate the parameters $(\rho_0, \sigma, \alpha)$ of our skewed Gaussian-inspired dark matter model to the astrophysically motivated parameters $(\rho_{-2}, \beta)$ of the Einasto profile, thereby enabling a phenomenological bridge between our theoretical construction and observationally constrained dark matter halo models.

We further assume that the total dark matter mass is a correction to the black hole mass. Thus, the resulting spacetime metric incorporating the dark matter contribution takes the same form as Eq.\ \eqref{eq:metric_orig} but with $M$ replaced by the total mass parameter $M_\epsilon$, which consists of two parts: The intrinsic (constant) black hole mass $M_{\rm BH}$ and a spatially varying dark matter correction $M_{\rm DM}(r)$,
\begin{equation}
M_\epsilon = M_{\rm BH} + M_{\rm DM}(r),
\end{equation}
where $M_{\rm DM}$ is related to the mass density $\rho_{\rm DM}(r)$ via the integral,  
\begin{equation}
    M_{\rm DM} = 
    \begin{cases}
        4\pp \int_{r_{\rm ISCO}}^r \dif x\, x^2 \rho_{\rm DM}(x), & r\ge r_{\rm ISCO}\\
        0, & r<r_{\rm ISCO}
    \end{cases}  
\end{equation}
as given by Eq.~\eqref{eq:dm_mass}.
It should be emphasized that our analysis addresses only the clustering of dark matter around regular black holes, 
consistent with recent findings based on binary companion orbital decay, such as Ref.\ \cite{Chan:2022gqd,Belikov:2013nca}. 
We do not consider scenarios in which dark matter itself emits gravitational waves spontaneously 
(e.g., via ultralight boson annihilation forming scalar clouds), as examined in e.g.\ the LIGO–Virgo–KAGRA Collaboration’s search for long lasting and quasi-monochromatic signals\cite{LIGOScientific:2021rnv}. 
Rather, our focus is strictly on how the ringdown-phase gravitational wave signature is altered when a regular black hole is embedded in a dark matter environment.

We now turn to the QNM analysis of the modified black hole with dark matter. To avoid unnecessary complexity, we focus on two representative cases.
As discussed in the previous section, the original (unmodified) model exhibits two distinct types of effective potentials: Type I
only contains a single potential barrier, and Type II contains both a barrier and an additional potential well, see the blue, orange and purple curves for Type I and the green curve for Type II in Fig.\ \ref{fig:Veff_orig}.
Given the structure introduced by the skewed dark matter distribution, we select two illustrative perturbation types for a detailed analysis: Scalar field perturbation and positive-parity spinor field perturbation, which 
correspond to the two characteristic types of effective potentials, respectively.

\subsection{Quasinormal mode analysis}

We now analyze how the introduction of dark matter affects the ringdown waveform, focusing particularly  on the case where the peak position of the dark matter distribution, $r_0$, lies outside the ISCO radius, $r_{\rm ISCO}$. Since $r_{\rm ISCO}$ necessarily lies outside the photon sphere radius $r_{\rm PS}$, and $r_{\rm PS}$ typically corresponds to the peak of the unmodified effective potential, especially in the eikonal limit, the dark matter distribution is effectively concentrated to the right of the peak of the unmodified  effective potential.

\begin{figure}[!ht]
	\centering
	\begin{subfigure}[b]{0.45\textwidth}
		\centering
			\includegraphics[width=\textwidth]{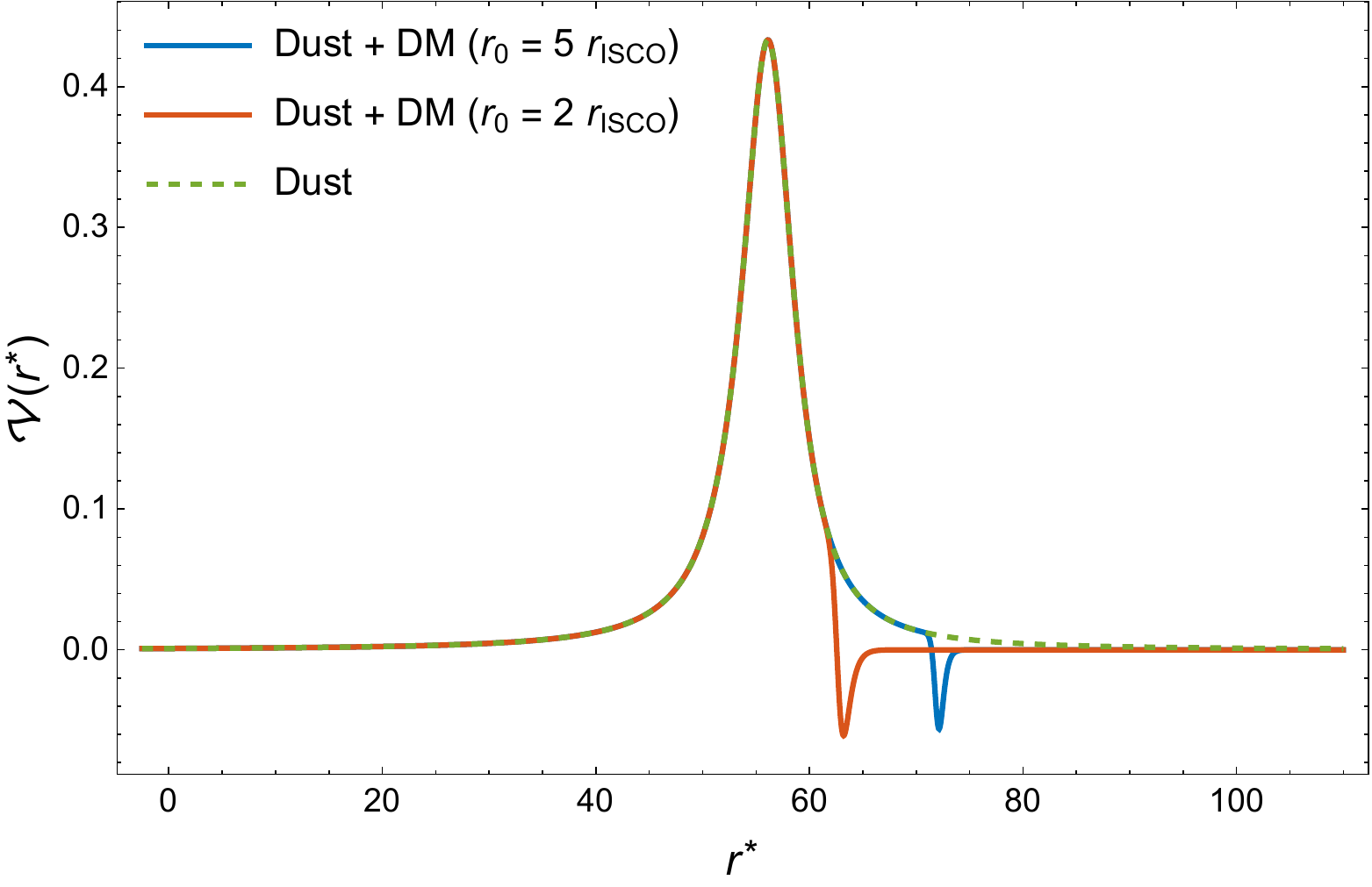}
	\caption{$(s)$}
	\label{fig:Veff_modi_s0}
	\end{subfigure}
	\begin{subfigure}[b]{0.45\textwidth}
		\centering
			\includegraphics[width=\textwidth]{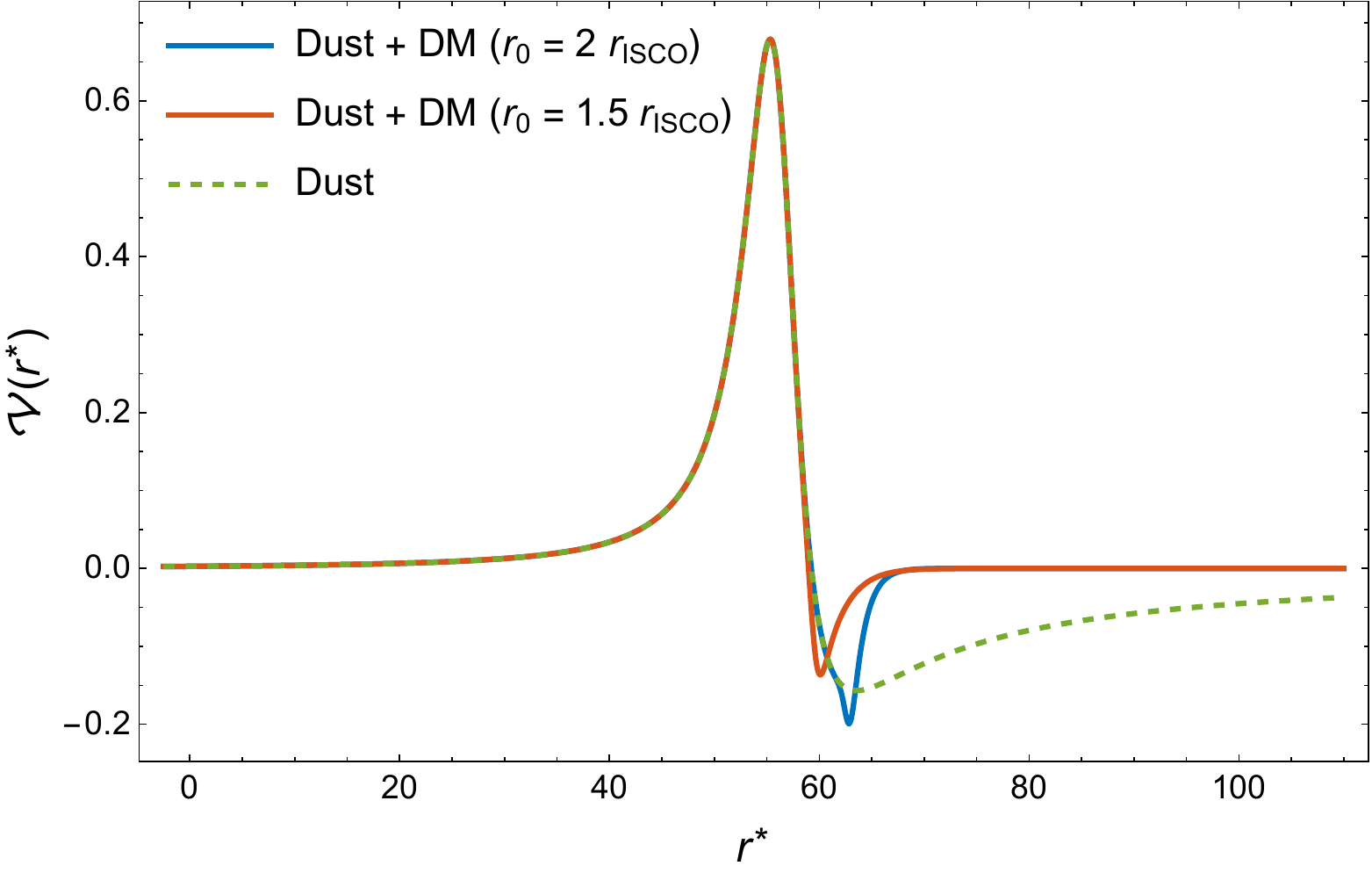}
	\caption{$(d,+)$}
	\label{fig:Veff_modi_sp}
	\end{subfigure}
	\captionsetup{width=.9\textwidth}
	\caption{Effective potentials of scalar and spinor field perturbations. The parameters are set to be $\rho_0=0.1$, $\sigma=4$, $\alpha=10$, $M=0.5$, $\xi=\xi_c$ and $\ell = 1$.}
	\label{fig:Veff_modi}
\end{figure}

The effective potentials for scalar field perturbations and positive-parity spinor field perturbations are shown in Fig.~\ref{fig:Veff_modi}, where the green dashed curves represent the unmodified potential from the dust collapse model, but the orange and blue solid curves include the corrections due to dark matter, where we set $r_0 = 2\, r_{\rm ISCO}$ for the orange curve and $r_0 = 5\, r_{\rm ISCO}$ for the blue curve in Type I and $r_0 = 1.5\, r_{\rm ISCO}$ for the orange curve  and $r_0 = 2\, r_{\rm ISCO}$ for the blue curve in Type II, respectively. As expected, the two modified potentials coincide to the unmodified ones on the left side of the primary peak, indicating that the dark matter has negligible influence in this region. However, the significant deviations appear on the right side of the peak.

In the scalar field case Fig.~\ref{fig:Veff_modi_s0}, i.e., Type I, the dark matter correction induces a shallow potential well on the right side of the peak. In contrast, for the spinor field perturbation Fig.~\ref{fig:Veff_modi_sp}, i.e., Type II, although the modification also occurs on the right side of the main peak, the effect is structurally different: The existing potential well becomes narrower in width (the orange and blue wells are narrower than the dashed green well), indicating a localized compression of the potential structure due to the presence of dark matter.

Thus, the primary effect of introducing a skewed normal correction to the mass profile is the formation of a relatively small potential well in the Type I effective potential, and it gives rise to a narrower well in the Type II effective potential. This behavior is qualitatively different from that studied in Ref.\ \cite{Cheung:2021bol,Berti:2022xfj}, where a direct Gaussian modification to the effective potential produced a barrier-like correction instead.

\begin{figure}[!ht]
	\centering
	\begin{subfigure}[b]{0.45\textwidth}
		\centering
			\includegraphics[width=\textwidth]{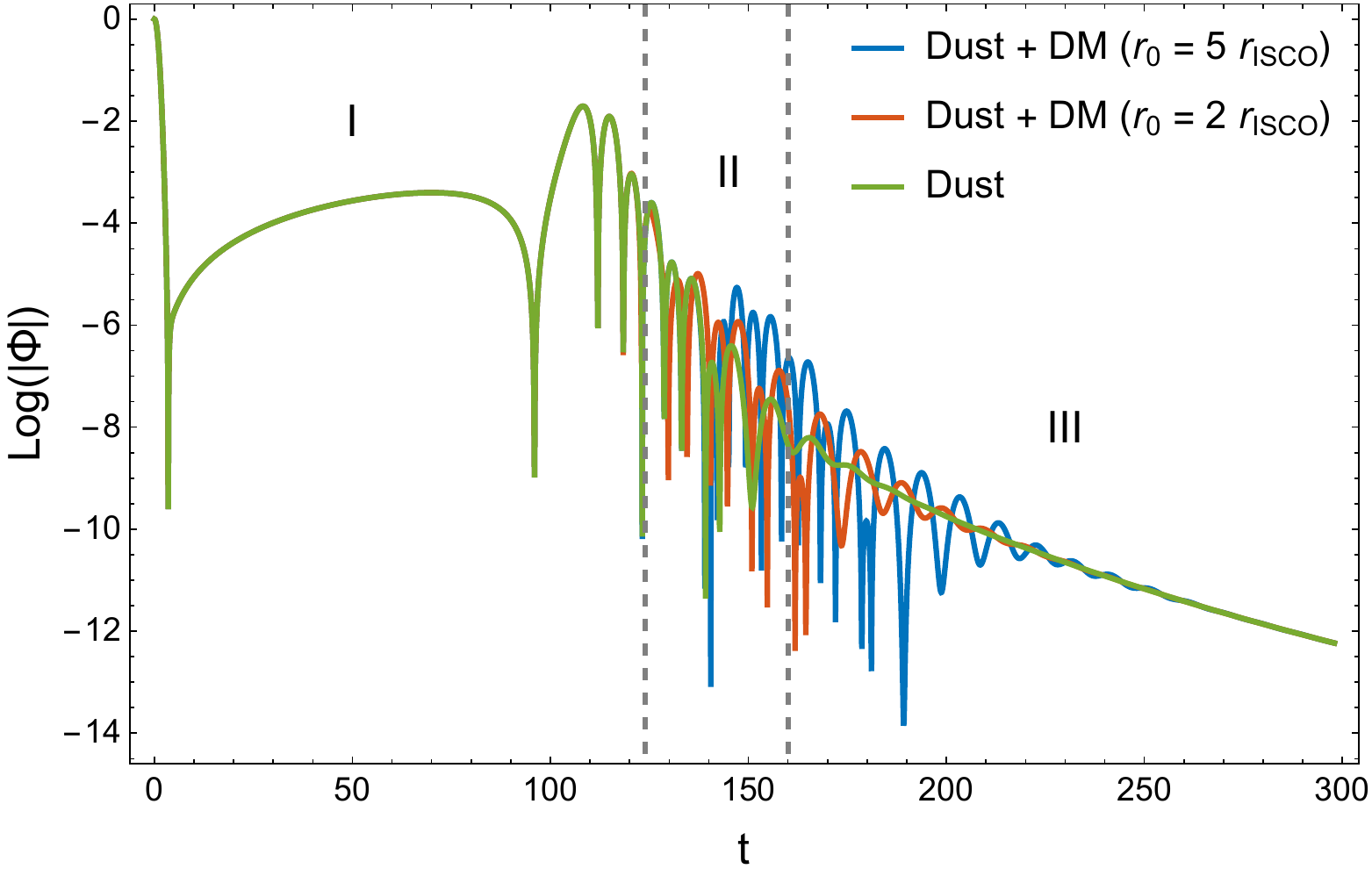}
	\caption{$(s)$}
	\label{fig:wf_modi_s0}
	\end{subfigure}
	\begin{subfigure}[b]{0.45\textwidth}
		\centering
			\includegraphics[width=\textwidth]{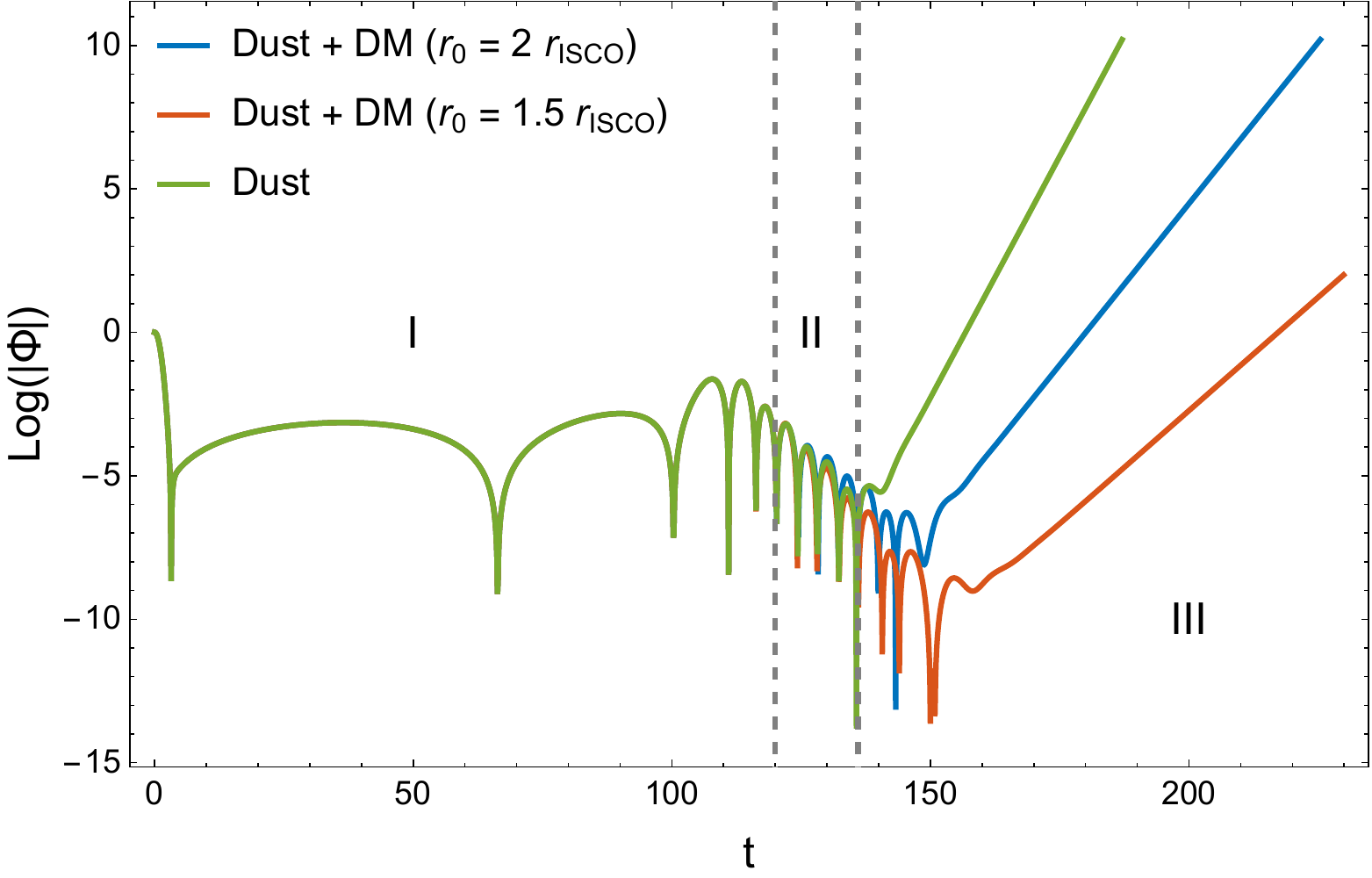}
	\caption{$(d,+)$}
	\label{fig:wf_modi_sp}
	\end{subfigure}
	\captionsetup{width=.9\textwidth}
	\caption{Waveforms of scalar and spin field perturbations corrected by dark matter in diagram \ref{fig:wf_modi_s0} and diagram \ref{fig:wf_modi_sp}, respectively. 
	The parameters are set  to be $\rho_0=0.1$, $\sigma=4$, $\alpha=10$, $M=0.5$, and $\xi=\xi_c$.}
	\label{fig:wf_modi}
\end{figure}

The ringdown waveforms corresponding to the two Types of effective potentials are presented in Fig.~\ref{fig:wf_modi}.
It is evident that the introduction of dark matter leads to distinct waveform features, even under identical temporal evolution conditions.

For scalar field perturbations Fig.~\ref{fig:wf_modi_s0}, 
it is evident that in Region I, the three waveforms overlap almost perfectly, showing no discernible differences. However, significant differences emerge in the subsequent two regions. In Region II, the amplitude exhibits varying degrees of enhancement, where the degree of amplification increases as $r_0$ becomes large.
In Region III, the oscillation in the unmodified model (the green curve) decays largely and vanishes soon, whereas in the dark matter-modified model, a persistent oscillatory behavior is maintained for a much longer duration. This implies that the waveform does not exhibit instability (such as secular growth or divergence over time) even when the abundance of dark matter is relatively small. Moreover, as the central position of the dark matter distribution shifts further outward (to a larger $r_0$), the oscillations become increasingly long-lived.
In summary, the oscillations possess a sustainable and long-lived feature, highlighting a potentially observable signature of surrounding dark matter distributions.

For spinor field perturbations Fig.~\ref{fig:wf_modi_sp}, the behavior is slightly different. At first, we did not adopt the same parameters as those in scalar field perturbations but smaller because we found that the waveform of the modified model almost overlaps with that of the unmodified model when $r_0$ is placed far from the main peak of the effective potential, say $r_0 = 5\,r_{\rm ISCO}$. This indicates that if the dark matter distribution is sufficiently far from black holes, its impact on the spinor field perturbation can be safely neglected.
Secondly, for dark matter distributed outside the ISCO, the amplitude in Region II is not necessarily enhanced compared to the unmodified waveform. When the center of the dark matter distribution is close to black holes, the amplitude tends to decrease slightly (see the orange curve); conversely, when the center is further away, the amplitude shows a slight increase (see the blue curve).
Finally, in Region III, the behavior of spinor field oscillations is completely different from that of scalar field oscillations, that is, the closer the center of the dark matter distribution is to black holes, the longer the oscillations persist. In particular, a late-time tail divergence appears in both the unmodified and modified models.

\section{Differential sensitivity of axial and polar tensor field perturbations}
\label{sec:tensorpert}

In this section, we turn our attention to tensor field perturbations. We demonstrate that axial and polar tensor 
perturbations have markedly different 
responses to dark matter modifications. 
Our analysis reveals that polar perturbations display significantly enhanced sensitivity compared to axial perturbations when subjected to identical dark matter distributions. This heightened sensitivity is sufficiently pronounced that even minimal dark matter concentrations can induce secondary peaks in the polar effective potential that substantially exceed the primary barrier height.

The effective potentials for axial $(a)$ and polar $(p)$ tensor field perturbations are given by
\cite{del-Corral:2022kbk}
\begin{subequations}
\begin{equation}
    \mathcal{V}^{(a)}=f(r) \left[
    -\frac{f'(r)}{r}+\frac{2 (f(r)-1)}{r^2}+\frac{\ell (\ell+1)}{r^2}
    \right],
\end{equation}
\begin{equation}
    \mathcal{V}^{(p)}=\frac{(\ell-1)^2 (\ell+2)^2 f(r)}{\lambda^2}
    \left[\frac{r^2 \beta^2 \left( r^2 \beta+3(\ell-1) (\ell+2)\right)}{3(\ell-1)^2 (\ell+2)^2}+\beta+\frac{(\ell-1) (\ell+2)+2}{r^2}\right],
\end{equation}
\end{subequations}
with
\begin{equation}
    \beta = \frac{f'(r)}{r}+\frac{2(1- f(r))}{r^2},\qquad
    \lambda=r f'(r)-2 f(r)+\ell^2+\ell.
\end{equation}

\begin{figure}[!ht]
	\centering
	\begin{subfigure}[b]{0.45\textwidth}
		\centering
		\includegraphics[width=\textwidth]{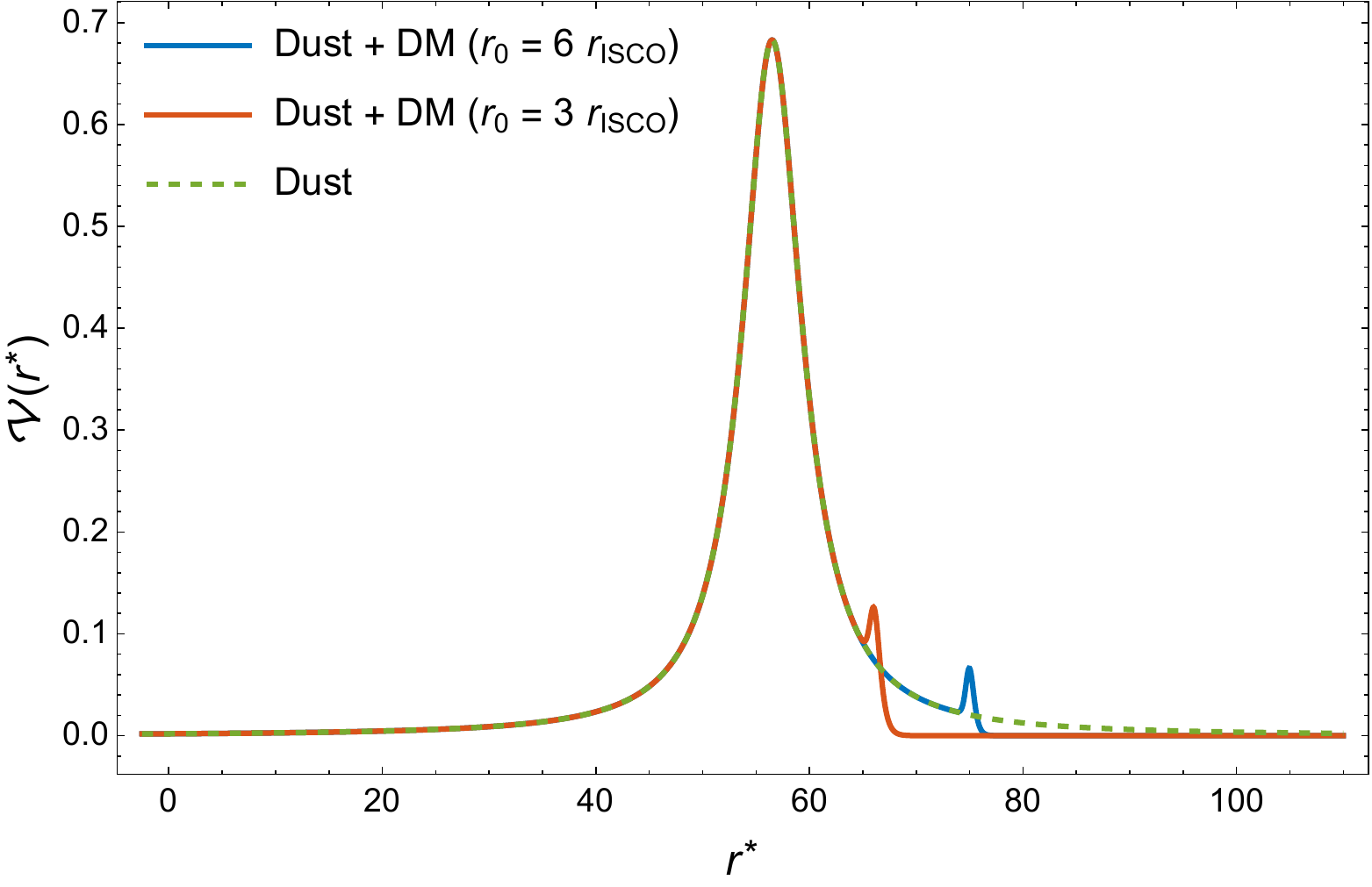}
		\caption{Axial}
		\label{fig:Veff_s2_axial}
	\end{subfigure}
	\begin{subfigure}[b]{0.45\textwidth}
		\centering
		\includegraphics[width=\textwidth]{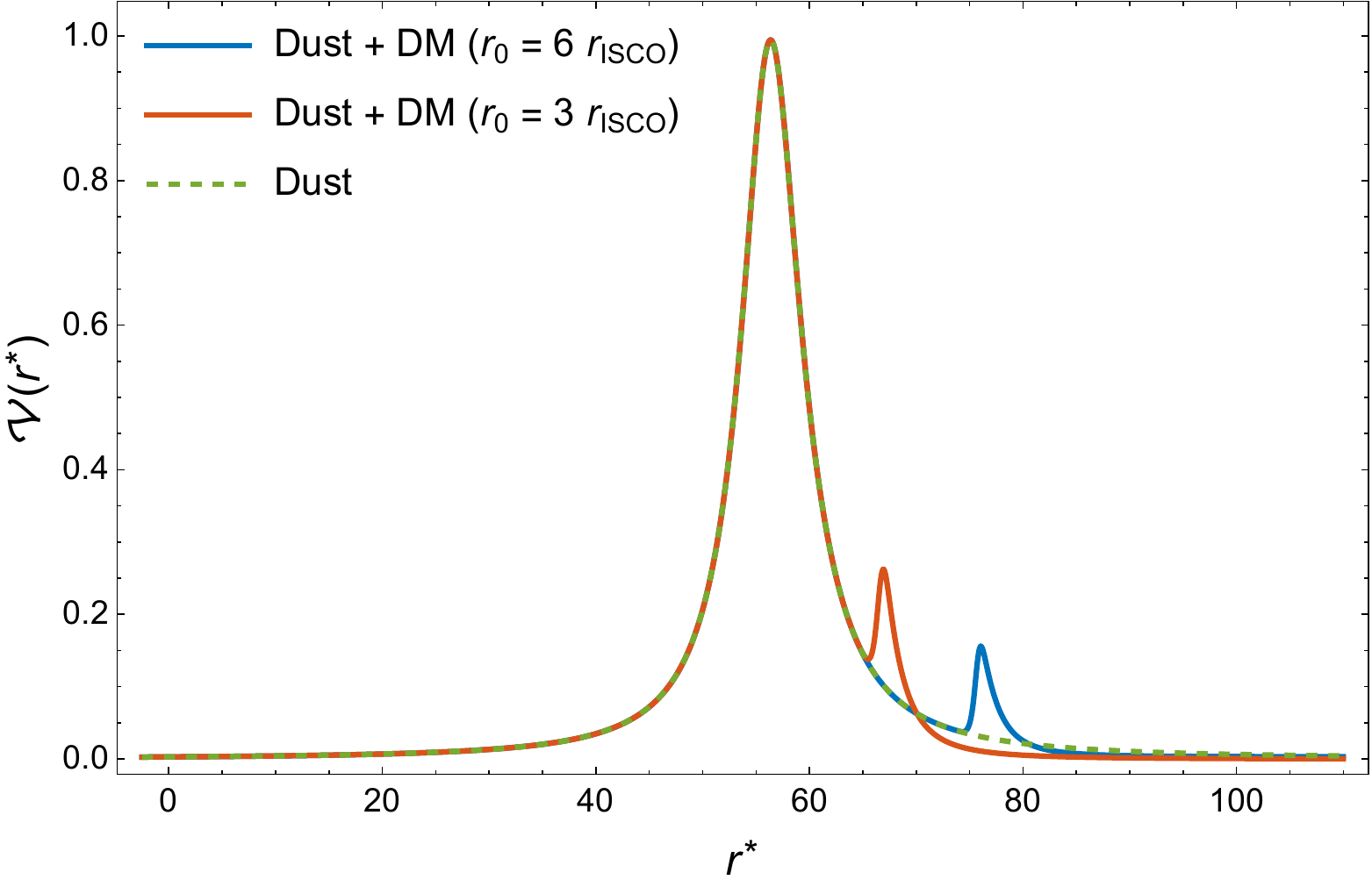}
		\caption{Polar}
		\label{fig:Veff_s2_polar}
	\end{subfigure}
	\captionsetup{width=.9\textwidth}
	\caption{Effective potentials of tensor field perturbations. The parameters are set to be $\sigma=4$, $\alpha=10$, $M=0.5$, $\xi=\xi_c$, and $\ell=2$. For axial perturbations, $\rho_0=0.1$ is unchanged for the two different values of $r_0$. For polar perturbations, $\rho_0=0.007$ at $r_0=3\,r_{\rm ISCO}$ and $\rho_0=0.002$ at $r_0=6\,r_{\rm ISCO}$ are chosen, respectively, demonstrating the enhanced sensitivity of polar perturbations.}
	\label{fig:veff_s2_modi}
\end{figure}

\begin{figure}[!ht]
	\centering
	\begin{subfigure}[b]{0.45\textwidth}
		\centering
		\includegraphics[width=\textwidth]{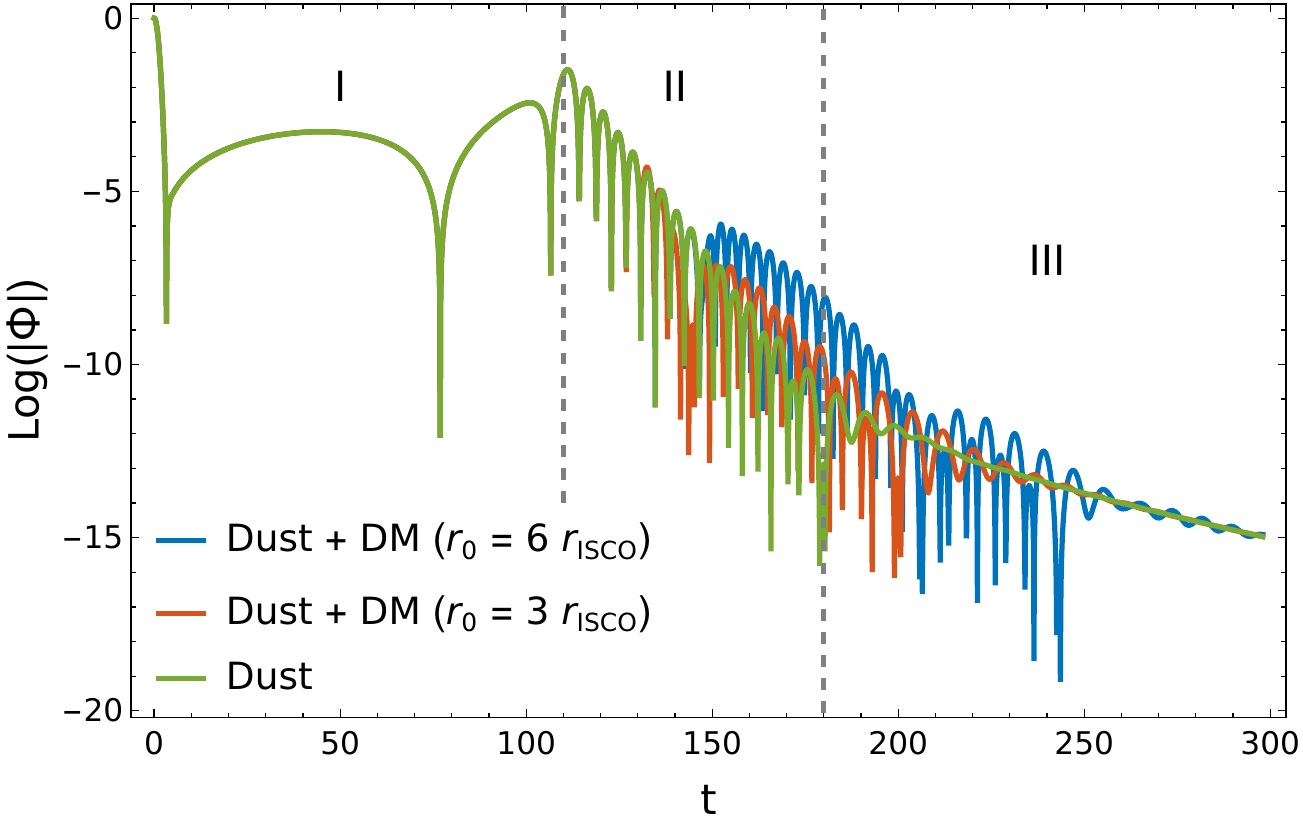}
		\caption{Axial}
		\label{fig:wf_dust_modi_axial}
	\end{subfigure}
	\begin{subfigure}[b]{0.45\textwidth}
		\centering
		\includegraphics[width=\textwidth]{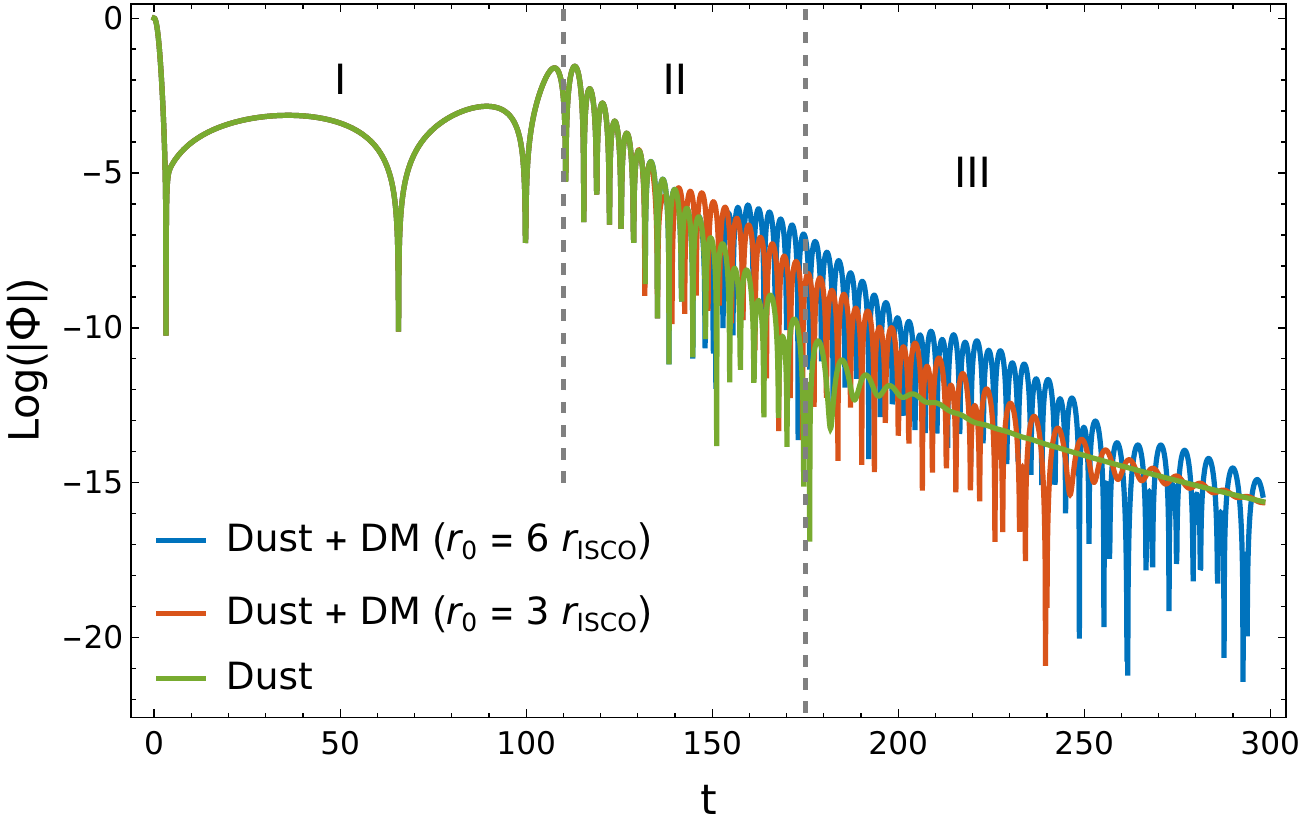}
		\caption{Polar}
		\label{fig:wf_dust_modi_polar}
	\end{subfigure}
	\captionsetup{width=.9\textwidth}
	\caption{Ringdown waveform for tensor field perturbations, where the parameters are set as in Fig.~\ref{fig:veff_s2_modi}. Despite requiring significantly lower dark matter abundances, polar perturbations generate more pronounced echo signatures with extended oscillation periods.}
	\label{fig:wf_s2_modi}
\end{figure}

We begin by considering the case without dark matter. 
For the metric  originated from dust collapse, the effective potentials for both axial and polar tensor field perturbations exhibit the identical form, as shown by the dashed green curves in Fig.~\ref{fig:veff_s2_modi}. 
This structural similarity naturally leads to the equivalent ringdown waveform for the both types of gravitational field perturbations, as illustrated by the green curves in Fig.~\ref{fig:wf_s2_modi}.
We select the quadrupole mode ($\ell=2$) for gravitational field perturbations, the lowest-order multipole producing gravitational waves in general relativity. Monopole ($\ell=0$) and dipole ($\ell=1$) modes are absent due to the energy-momentum and momentum conservation, where the two modes imply the time-varying total mass or net center-of-mass acceleration. Thus, $\ell=2$ is the physically relevant starting mode for analyzing gravitational wave emission.

We now incorporate dark matter into our analysis. Following the approach in Ref.~\cite{Figueiredo:2023gas}, we neglect the backreaction of the energy-momentum tensor in the perturbation equations. This approximation simplifies the mathematical treatment but remains valid under the assumption of a small density parameter $\rho_0$, which ensures that higher-order corrections from the stress-energy tensor are suppressed. Although this neglect may introduce systematic errors in the effective potential, particularly for configurations with dense dark matter distributions, the resulting framework provides a tractable description of the system's dynamics.

We consider a skewed normal distribution for the dark matter density profile, as specified in Eq.~\eqref{eq:dm_mass}. The corresponding effective potentials are presented in Figs.~\ref{fig:Veff_s2_axial} and \ref{fig:Veff_s2_polar} for axial and polar perturbations, respectively. As we shall demonstrate below, the presence of dark matter breaks the symmetry between axial and polar modes that characterizes the vacuum case, leading to qualitatively different perturbation dynamics.
A crucial observation emerges from our analysis: The introduction of dark matter produces barrier modifications rather than potential well modifications, contrasting sharply with the test field scenarios examined in the previous section.
This modification to the main potential barrier is expected to induce echo-like features in the resulting waveform. Such a behavior can be anticipated from the emergence of a double-peak structure in the effective potential. These configurations act effectively as cavities, partially reflect and trap gravitational wave energy between successive potential barriers, and ultimately generate late-time echo signatures. This mechanism aligns with the previous analyses of gravitational wave echoes in modified gravity scenarios \cite{Huang:2021qwe}.

The differential sensitivity between axial and polar perturbation modes is particularly striking. Figure~\ref{fig:Veff_s2_axial} shows axial perturbations with uniform parameter $\rho_0=0.1$, while Fig.~\ref{fig:Veff_s2_polar} demonstrates that polar perturbations achieve comparable potential modifications with $\rho_0$ values two orders of magnitude smaller: $\rho_0=0.007$ for $r_0=3\,r_{\rm ISCO}$ and $\rho_0=0.002$ for $r_0=6\,r_{\rm ISCO}$. This dramatic difference in required dark matter abundance underscores the exceptional sensitivity of polar perturbations and suggests that they may serve as more efficient probes of dark matter signatures in gravitational wave observations.

The corresponding waveforms for axial and polar perturbations are displayed in Figs.~\ref{fig:wf_dust_modi_axial} and \ref{fig:wf_dust_modi_polar}, respectively, both exhibiting distinct echo phenomena with varying amplitudes and durations.

The comparative analysis of the waveforms reveals several universal features: (i) the modified waveforms in Region I remain virtually indistinguishable from the dark matter-free case; (ii) a greater separation between the dark matter center $r_0$ and the event horizon correlates with higher first-echo amplitudes and extended oscillation tails; and (iii) polar perturbations consistently produce more pronounced echoes with longer-duration oscillations although the dark matter abundance is two orders of magnitude lower than that of axial perturbations. These findings suggest polar tensor field perturbations as particularly sensitive probes for detecting dark matter signatures in gravitational wave astronomy.

\section{Discussion and outlook}
\label{sec:dis}
In this work, we have investigated the ringdown waveform of a class of regular black holes formed via gravitational collapse with quantum corrections, specifically focusing on how skewed dark matter distributions modify gravitational wave signatures. By introducing dark matter in the form of a spatially asymmetric (skewed normal) distribution centered outside the ISCO, we have derived self-consistent modifications to the background metric, and consequently to the effective potential governing various field  perturbations.

Our analysis demonstrates that dark matter, even with a modest abundance, can lead to significant structural changes in the effective potential, typically inducing a shallow potential well for scalar and spinor field perturbations, and occasionally introducing a secondary barrier structure in the axial and polar sectors. These modifications result in qualitatively distinct ringdown waveforms, including long-lived modes and altered decay rates. Interestingly, our findings also reveal that the same dark matter distribution yields different sensitive deformations depending on the type of perturbations, highlighting the intrinsic asymmetry between axial and polar gravitational modes in such spacetimes.

Looking forward, our analyses suggest several interesting issues. At first, a full treatment including the backreaction of energy-momentum tensors could further refine the stability analysis and provide more accurate waveform predictions. Then,  it would be essential for realistic astrophysical scenarios to extend our analyses to spinning regular black holes. At last, given that multi-barrier potentials may lead to gravitational wave echoes under suitable conditions, a future work may explore whether tuned dark matter configurations around regular black holes could produce observable echo signals, providing a potential new probe of quantum gravity effects.

Finally, as gravitational wave detectors improve in precision, it becomes increasingly tangible for us to detect deviations from singular black hole signatures due to surrounding dark matter distributions or the absence of a central singularity. Our study contributes to this growing effort by providing a theoretically consistent framework for modeling such deviations in the context of regular black holes surrounded by dark matter. One of urgent issues that needs to be solved is
to advance this framework toward more realistic astrophysical scenarios, i.e., to extend our analyses to rotating regular black holes. In such cases, the introduction of dark matter, especially with a skewed distribution, could interact nontrivially with the frame-dragging effects of rotation, potentially leading to qualitatively different modifications of ringdown waveforms. To model such systems, one promising idea is to employ the extended Newman–Janis algorithm \cite{Lan:2024wfo} or its generalizations to construct rotating counterparts of regular black holes with dark matter corrections. It will be a central focus of our future research to investigate the ringdown behavior in these rotating spacetimes and analyze the axial–polar mode splitting, damping times, and possible echoes.

\section*{Acknowledgements}

C.L.\ and Y.-L.\ T. make an equal contribution.
The authors thank Alexander Zhidenko for pointing out a typographical error in the formula.
This work was supported in part by the National Natural Science Foundation of China under Grant No.\ 12175108. Moreover, L.C. is supported by Yantai University under Grant No.\ WL22B224 and
Z.-X. Z is supported by the Pilot Scheme of Talent Training in Basic Sciences (Boling Class of Physics, Nankai University), Ministry of Education.

During the preparation of this work the authors used ChatGPT in order to create the schematic diagram shown in Figure 3. After using this tool/service, the authors reviewed and edited the content as needed and took full responsibility for the content of the published article.

\appendix

\section{Schwarzschild black hole surrounded by dark matter}
\label{app:schwarzschild}

In this appendix, we analyze the effect of introducing dark matter on the ringdown waveform of a Schwarzschild black hole by applying the identical mass modification as the above.
This comparison serves to validate our findings and explore the generality of dark matter effects across different black hole configurations.

We adopt such a parameter set: $M=1/2$, $\sigma=4$, and $\alpha=10$, examining two distinct central positions for the dark matter distribution. The dark matter abundance parameters are systematically varying to perturbation types in order to account for their differential sensitivities: For test field perturbations, $\rho_0=0.1$ and $\ell=1$; for axial gravitational field perturbations, $\rho_0=0.05$ and $\ell=2$; for polar  gravitational field perturbations, $\rho_0=0.0024$ and $\ell=2$ at $r_0=5\,r_{\rm ISCO}$, and $\rho_0=0.012$ and $\ell=2$ at $r_0=2\,r_{\rm ISCO}$, respectively. These parameter choices reflect the enhanced sensitivity of polar modes as demonstrated in our above analysis.

\begin{figure}[!ht]
	\centering
	\begin{subfigure}[b]{0.45\textwidth}
		\centering
		\includegraphics[width=\textwidth]{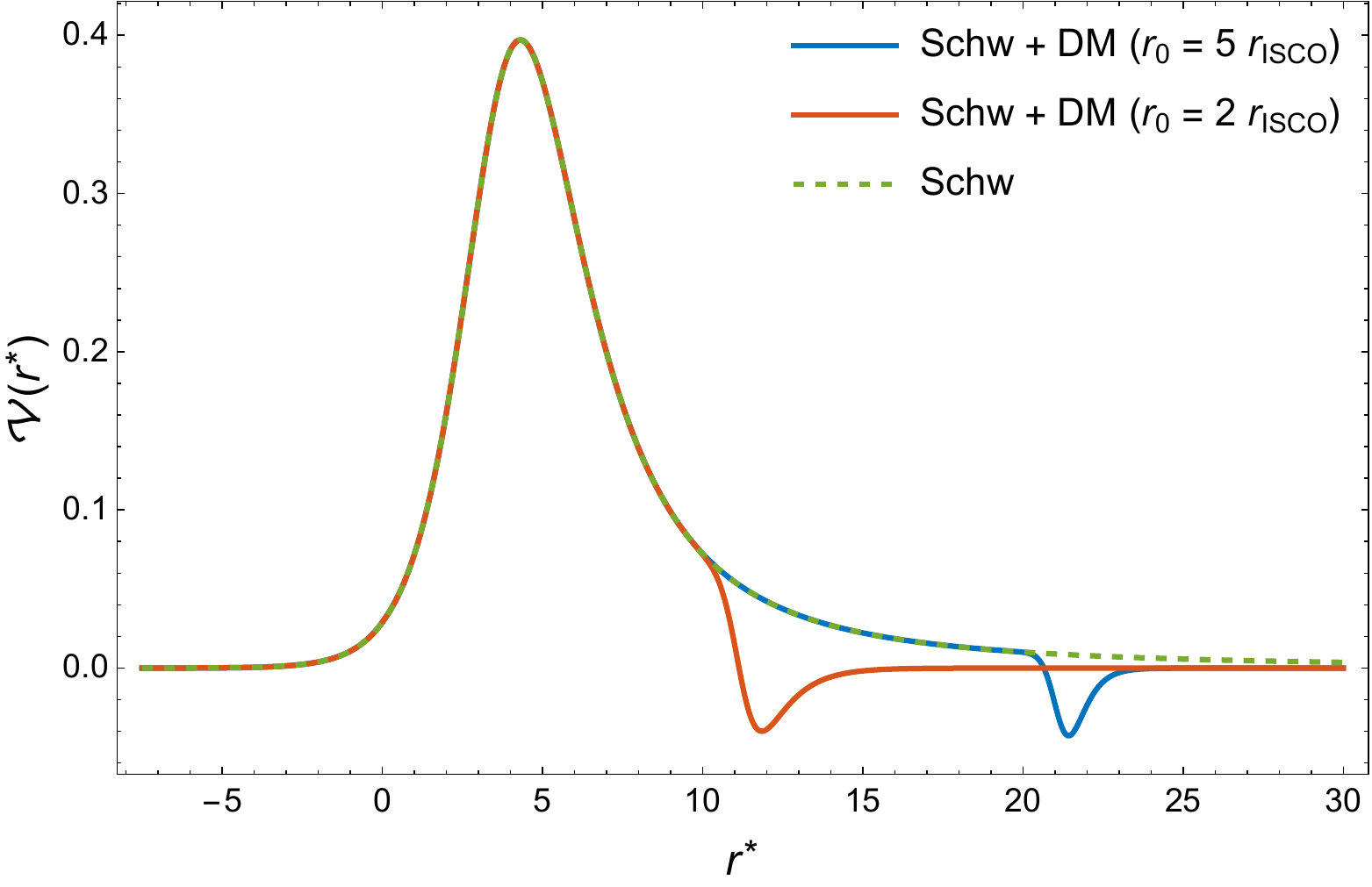}
		\caption{$(s)$}
		\label{fig:Veff_sch_modi_s0}
	\end{subfigure}
	\begin{subfigure}[b]{0.45\textwidth}
		\centering
		\includegraphics[width=\textwidth]{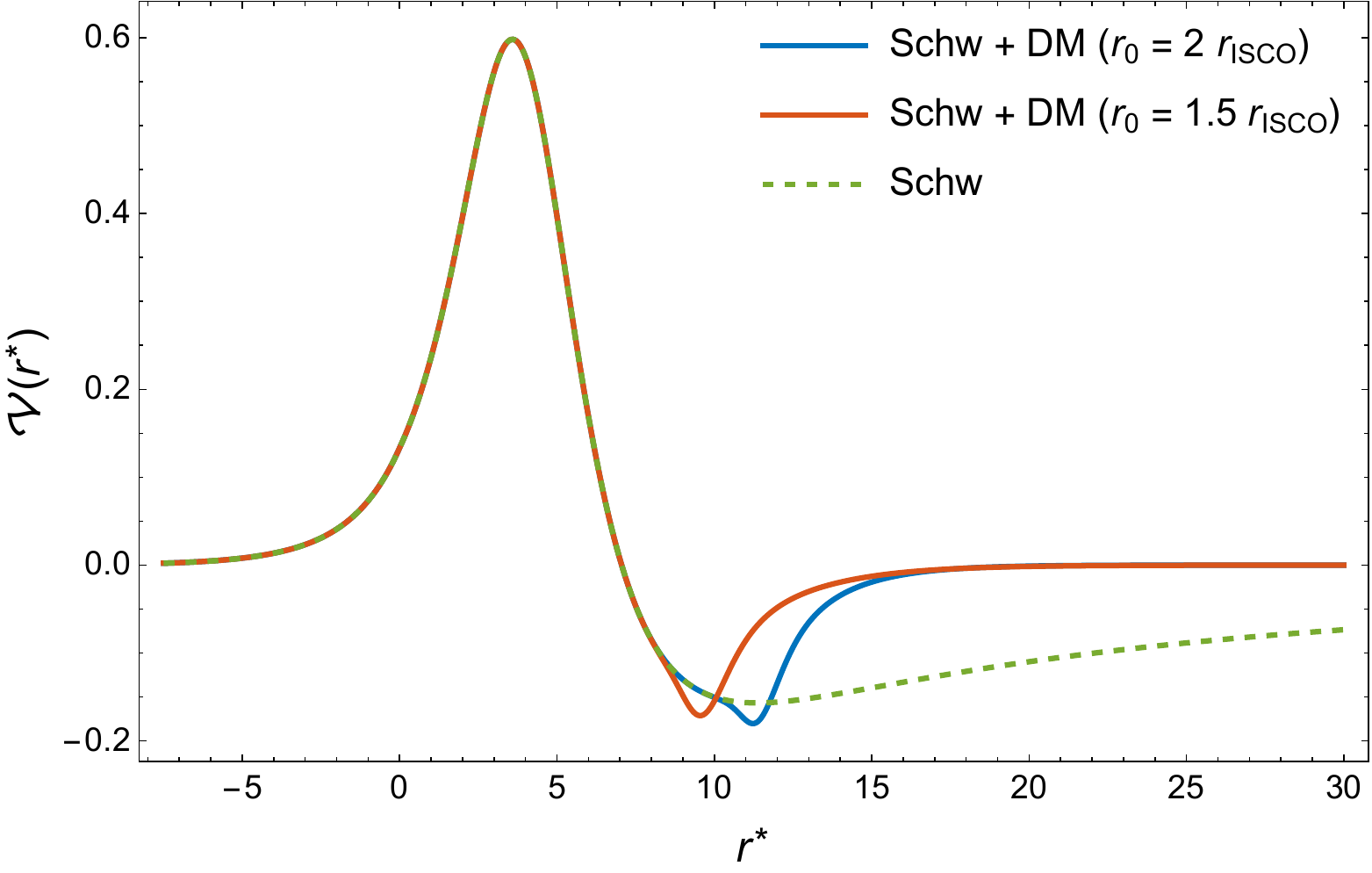}
		\caption{$(d,+)$}
		\label{fig:Veff_sch_modi_spin}
	\end{subfigure}
	\hfill
	\begin{subfigure}[b]{0.45\textwidth}
		\centering
		\includegraphics[width=\textwidth]{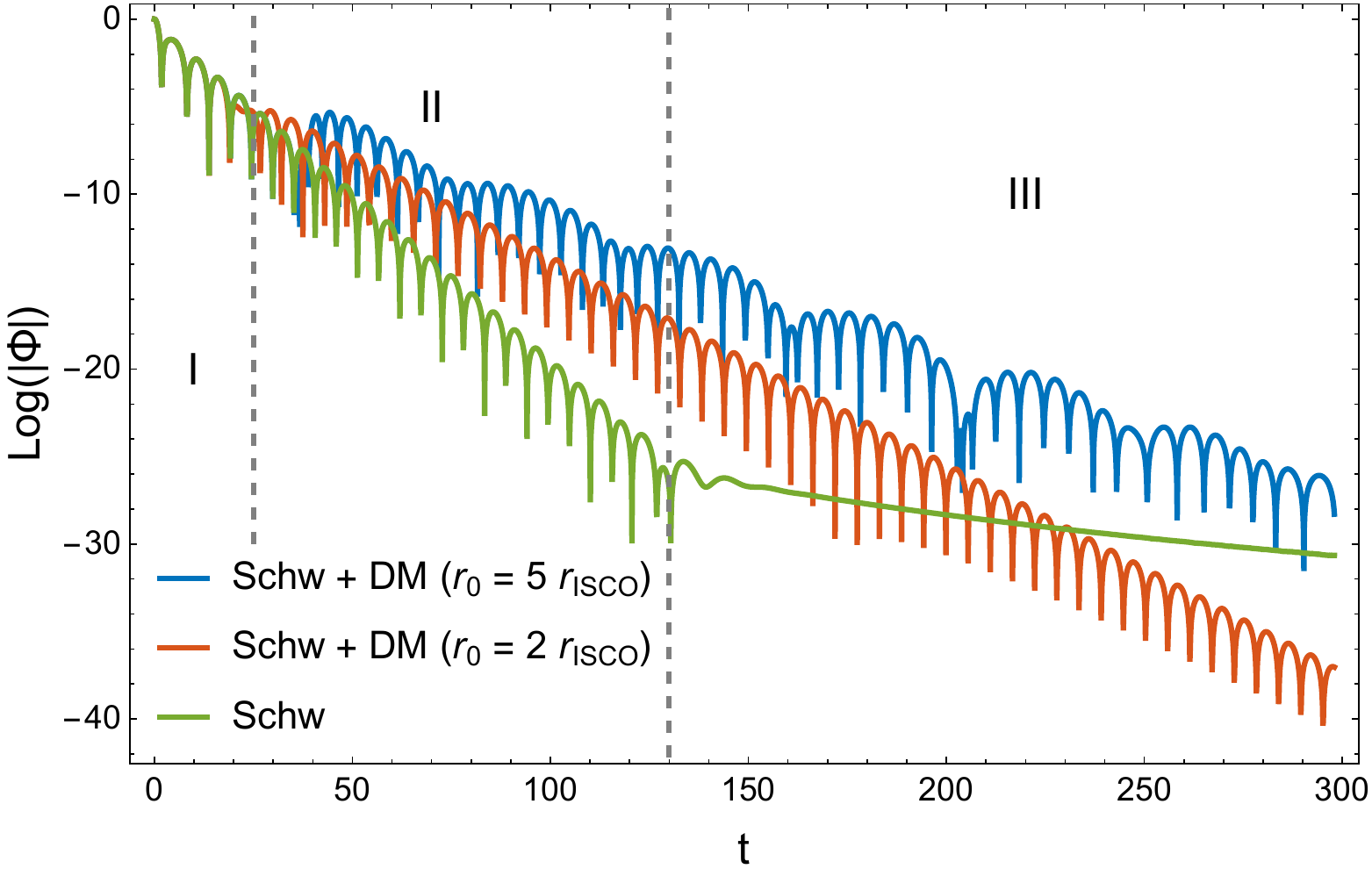}
		\caption{$(s)$}
		\label{fig:wf_sch_modi_s0}
	\end{subfigure}
	\begin{subfigure}[b]{0.45\textwidth}
		\centering
		\includegraphics[width=\textwidth]{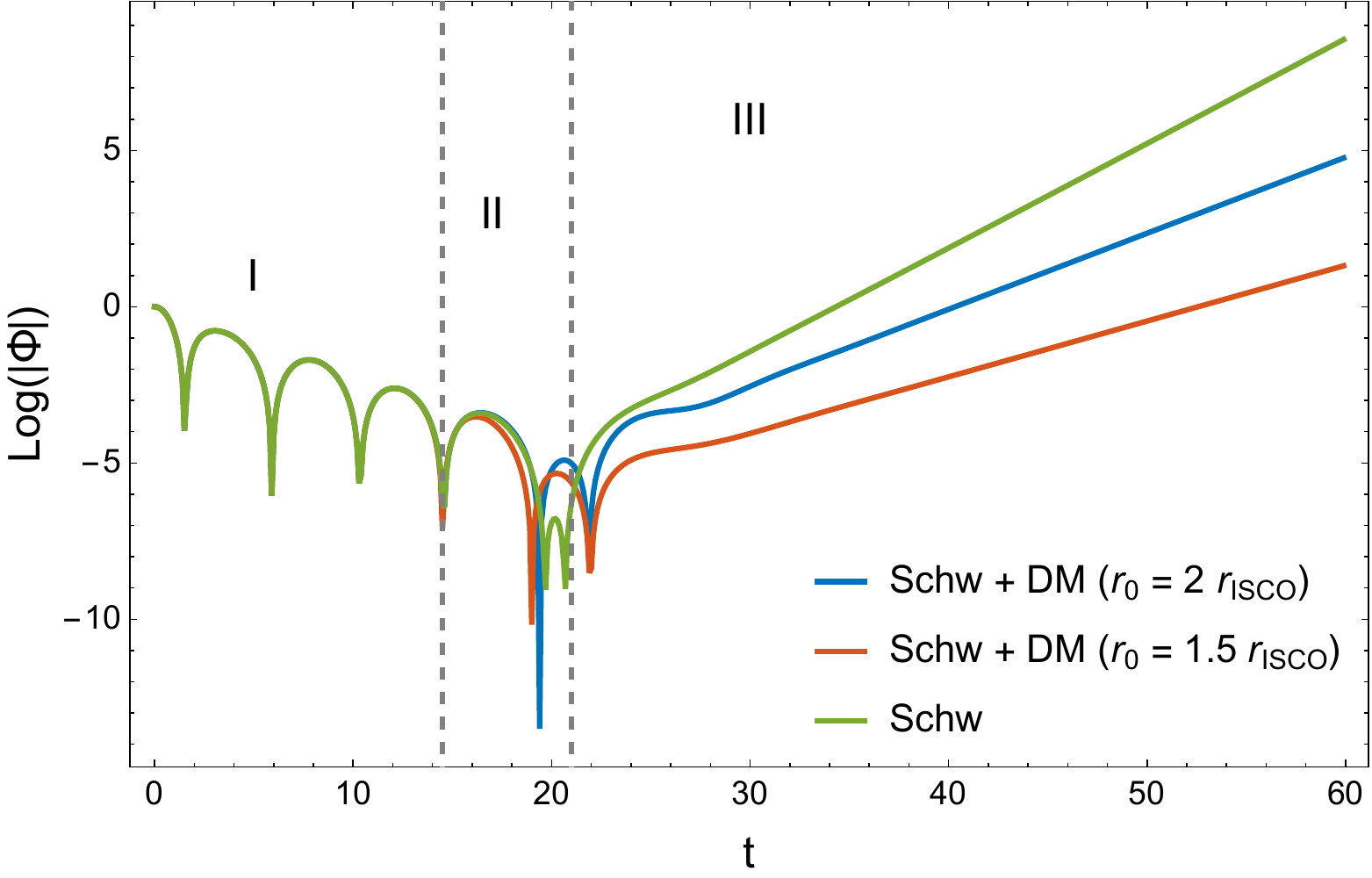}
		\caption{$(d,+)$}
		\label{fig:wf_sch_modi_spin}
	\end{subfigure}
	\captionsetup{width=.9\textwidth}
	\caption{Effective potentials and corresponding waveforms for scalar and spinor field perturbations of Schwarzschild black holes surrounded by dark matter ($\ell=1$). The dashed curves represent unmodified Schwarzschild potentials.}
	\label{fig:sch_modi_1}
\end{figure}

The effective potentials for scalar field perturbations and even-parity spinor field perturbations are shown in Figs.\ \ref{fig:Veff_sch_modi_s0} and \ref{fig:Veff_sch_modi_spin}, respectively. 
The dashed curves indicate the effective potential of Schwarzschild black holes. 
As seen from the plots, the effective potential predominantly coincides with Fig.~\ref{fig:Veff_modi}, supporting our earlier conclusion that a skewed normal distribution of dark matter primarily introduces a shallow potential well into the unmodified barrier structure.

The corresponding waveforms are displayed in Figs.\ \ref{fig:wf_sch_modi_s0} and \ref{fig:wf_sch_modi_spin}. Notably, for scalar field perturbations, the presence of dark matter gives rise to a pronounced long-lived mode. Moreover, as the dark matter center $r_0$ shifts from the left to the right, the waveform exhibits a slower decay, i.e., the absolute value of imaginary part of the QNM frequency decreases, as shown in Fig.\ \ref{fig:wf_sch_modi_s0}. This long-lived mode does not appear significantly in regular black holes.
In the case of the even-parity spinor field perturbation, the waveform exhibits a form similar to that discussed in the context of regular black holes. 
The introduction of dark matter leads to a weakening of the late-time tail divergence, and the closer $r_0$ is to Schwarzschild black holes, the weaker the degree of attenuation is, as seen in Fig.~\ref{fig:wf_sch_modi_spin}.

\begin{figure}[!ht]
	\centering
	\begin{subfigure}[b]{0.45\textwidth}
		\centering
		\includegraphics[width=\textwidth]{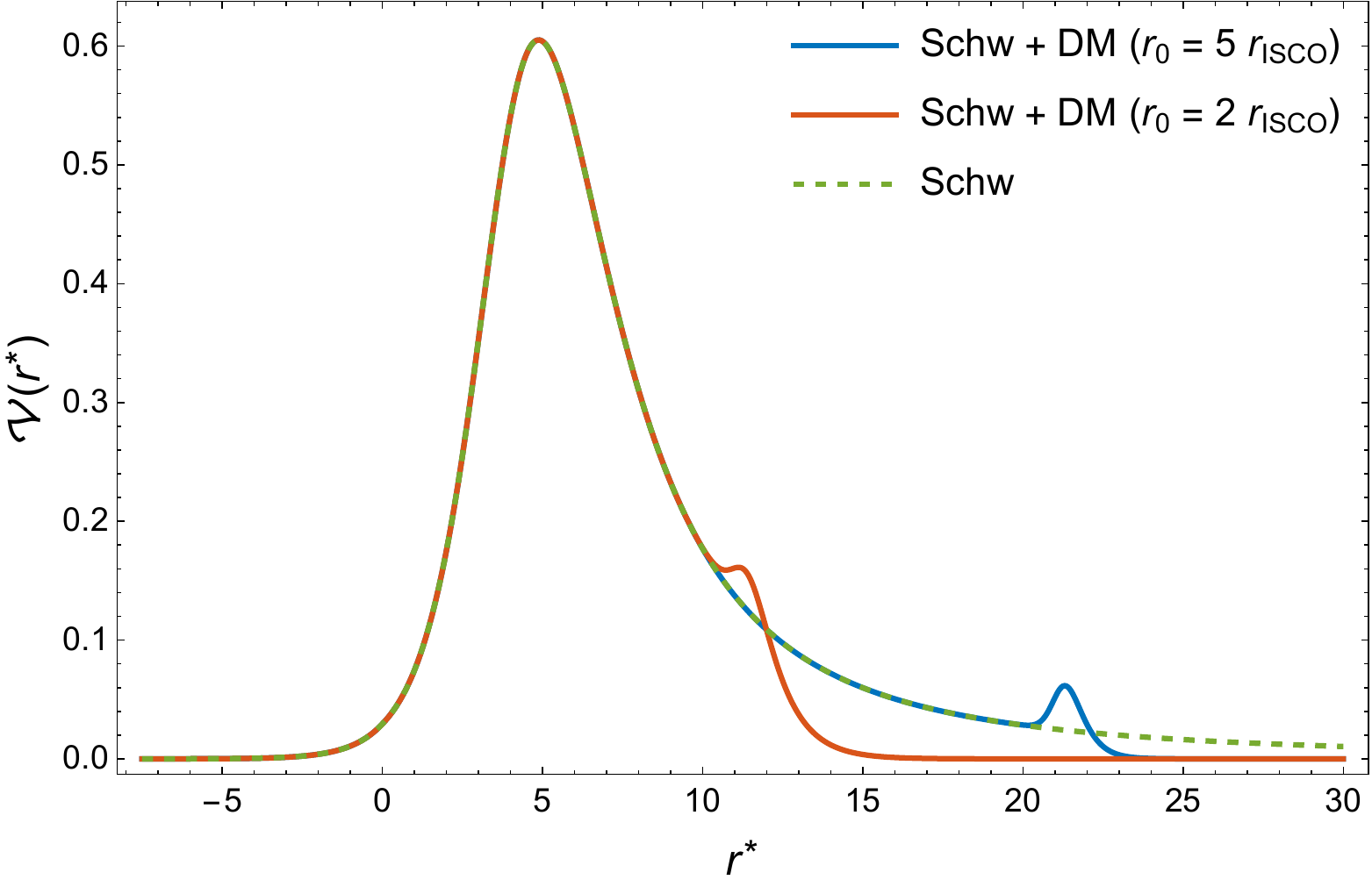}
		\caption{Axial}
		\label{fig:Veff_sch_modi_axial}
	\end{subfigure}
	\begin{subfigure}[b]{0.45\textwidth}
		\centering
		\includegraphics[width=\textwidth]{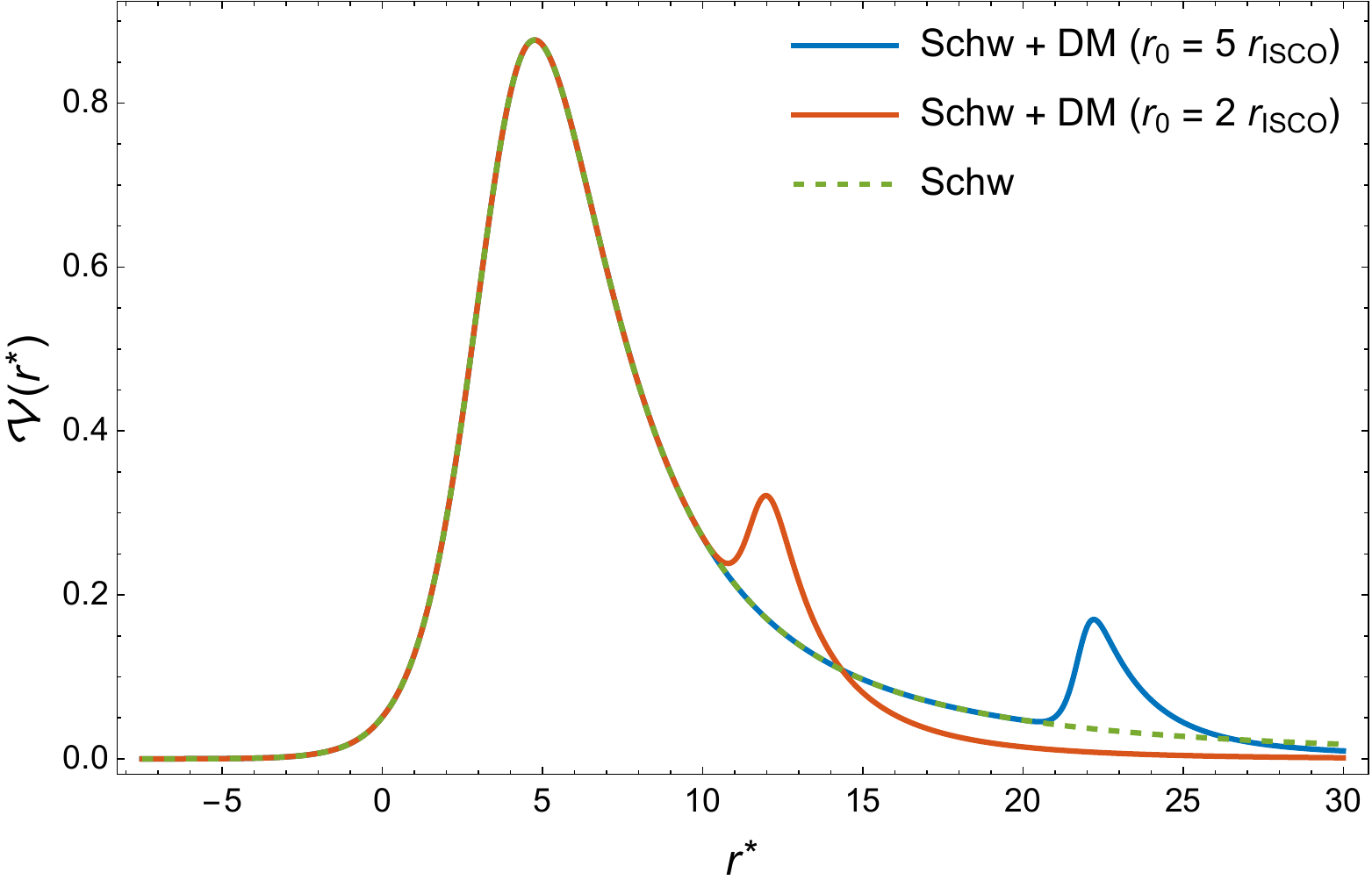}
		\caption{Polar}
		\label{fig:Veff_sch_modi_polar}
	\end{subfigure}
	\begin{subfigure}[b]{0.45\textwidth}
		\centering
		\includegraphics[width=\textwidth]{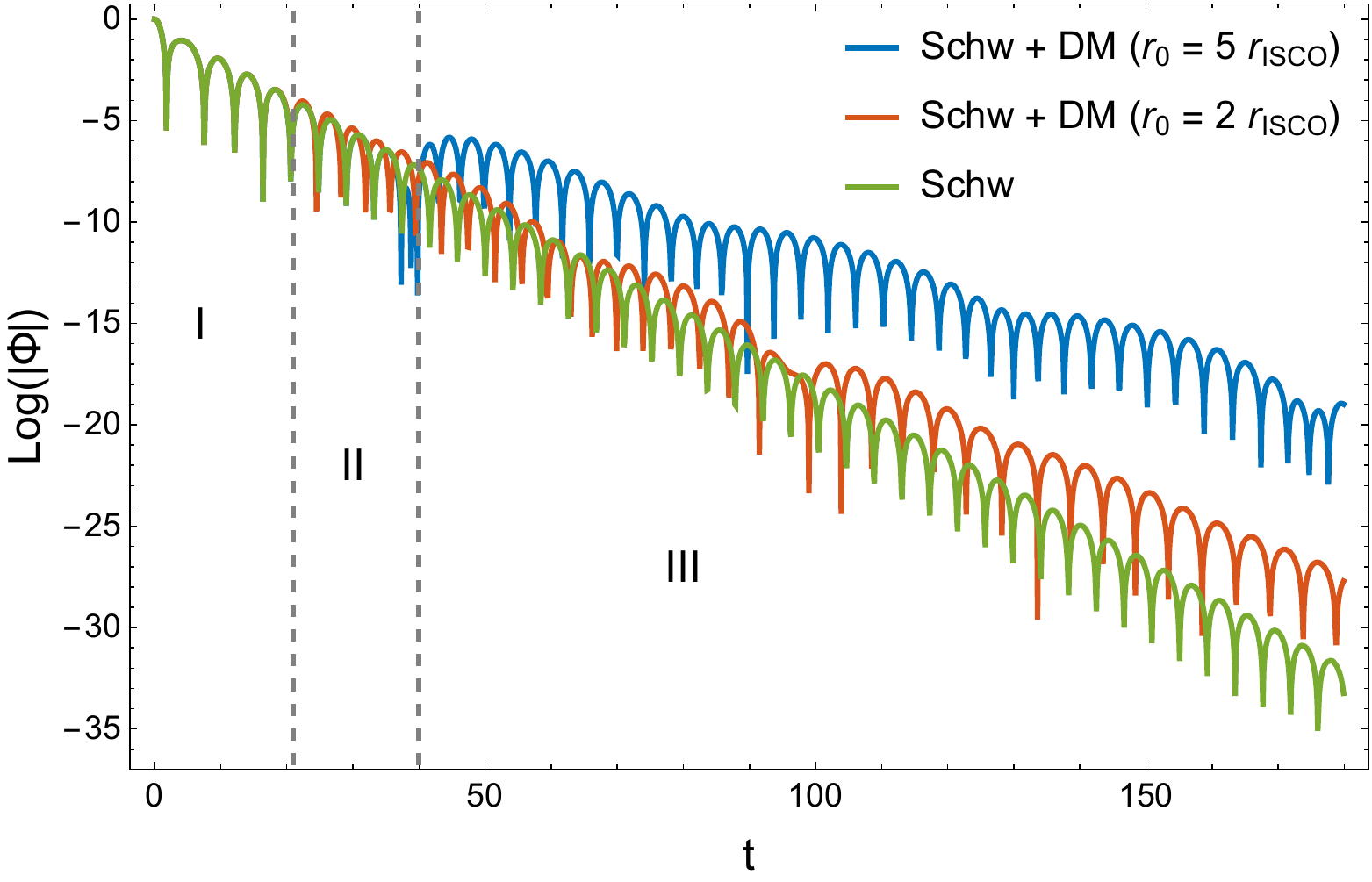}
		\caption{Axial}
		\label{fig:wf_sch_modi_axial}
	\end{subfigure}
	\begin{subfigure}[b]{0.45\textwidth}
		\centering
		\includegraphics[width=\textwidth]{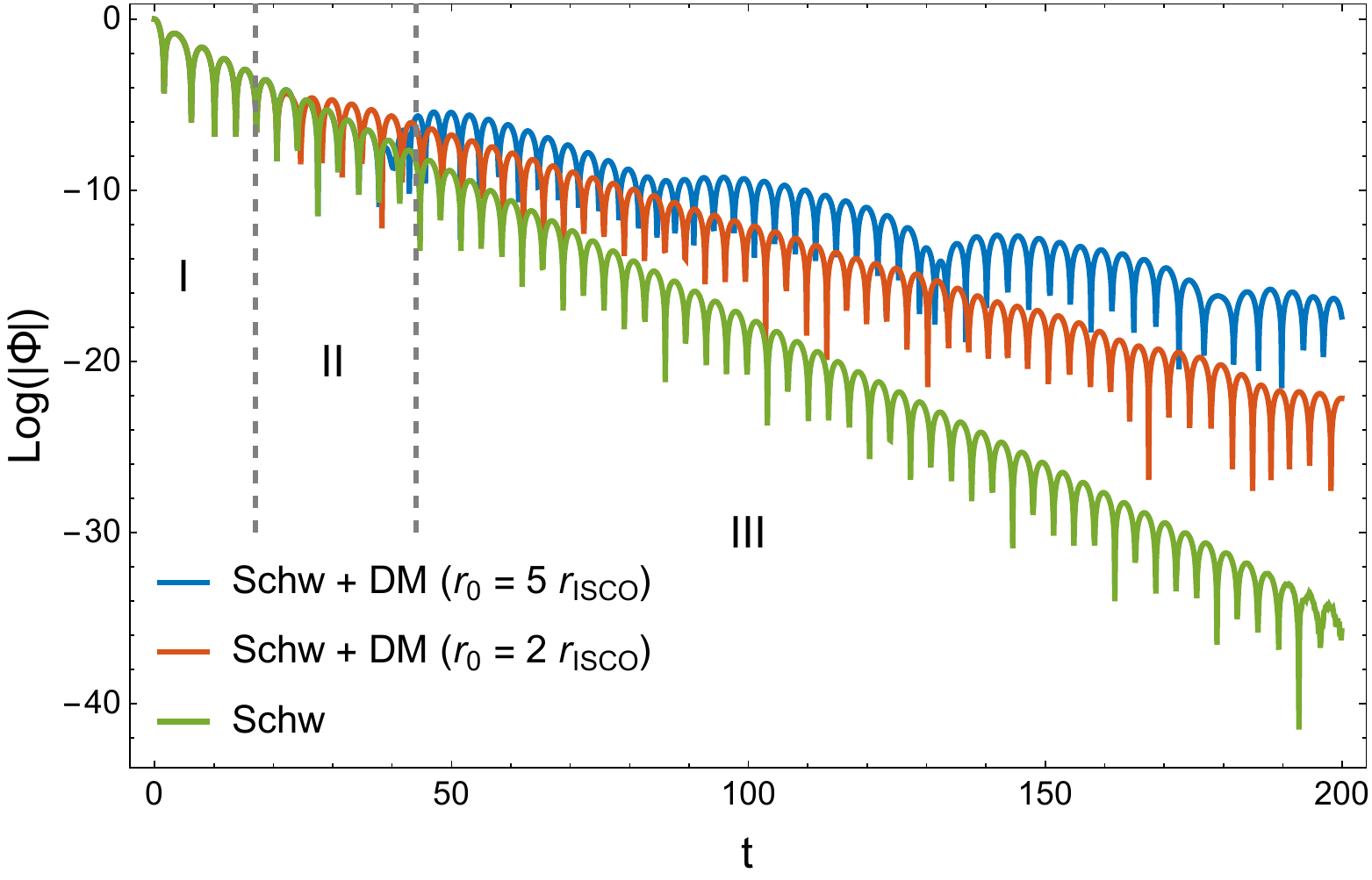}
		\caption{Polar}
		\label{fig:wf_sch_modi_polar}
	\end{subfigure}
	\captionsetup{width=.9\textwidth}
	\caption{Effective potentials and corresponding waveforms for axial and polar gravitational field perturbations of Schwarzschild black holes surrounded by dark matter ($\ell=2$). The enhanced sensitivity of polar perturbations produces more pronounced echo signatures despite lower dark matter abundances.}
	\label{fig:sch_modi_2}
\end{figure}

We now examine gravitational field perturbations, where the differential sensitivity between axial and polar modes becomes particularly pronounced. Figs.~\ref{fig:Veff_sch_modi_axial} and \ref{fig:Veff_sch_modi_polar} present the effective potentials for axial and polar perturbations, respectively. Crucially, these tensor field perturbations respond differently to dark matter compared to test field perturbations: Rather than introducing potential wells, dark matter generates additional barrier structures in the effective potential.

The enhanced sensitivity of polar perturbations to dark matter is clearly evident. Despite employing significantly lower dark matter abundances, polar perturbations exhibit more substantial potential modifications than their axial counterparts. This differential response arises from the distinct mathematical structures governing axial and polar gravitational field perturbations, as established in the main text.

When Gaussian-like mass corrections are introduced into the spacetime metric, the resulting differential sensitivities manifest as distinctly different waveform behaviors. This is demonstrated in Figs.~\ref{fig:wf_sch_modi_axial} and \ref{fig:wf_sch_modi_polar}, where polar perturbations exhibit significantly stronger echo phenomena compared to axial perturbations, consistent with our findings for regular black holes.

The emergence of double-barrier structures in the effective potentials provides the necessary conditions for gravitational wave echo generation \cite{Yang:2024prm}. However, the observation of distinct echo signals requires specific geometric constraints on the potential profile, particularly regarding the separation distance between successive barriers. In our configuration, while the barrier structures are present, the relatively modest separation distances suggest that clear echo identification would require highly sensitive gravitational wave detectors and sophisticated data analysis techniques.

Nevertheless, the systematic differences between axial and polar perturbations, combined with their differential sensitivity to dark matter, offer promising avenues for constraining dark matter properties through future gravitational wave observations. The enhanced response of polar modes, in particular, suggests that they may serve as particularly sensitive probes for detecting subtle dark matter signatures in the ringdown waveform.

\bibliographystyle{utphys}
\bibliography{references}
\end{document}